\documentclass[aps,prl,reprint,groupedaddress]{revtex4-1}

\usepackage{graphicx}
\usepackage{braket}
\usepackage{amsmath}
\usepackage{times}
\usepackage[english]{babel}

\newcommand{\nth}{n_{\mathrm{th}}}

\newcommand{\etain}{\eta_{\mathrm{in}}}

\newcommand{\nf}{n_{\mathrm{f}}}
\newcommand{\etadet}{\eta_{\mathrm{det}}}

\newcommand{\nba}{n_{\mathrm{ba}}}
\newcommand{\nimp}{n_{\mathrm{imp}}}
\newcommand{\etaeff}{\eta_{\mathrm{eff}}}
\newcommand{\nc}{n_{\mathrm{c}}}
\newcommand{\nadd}{n_{\mathrm{add}}}

\newcommand{\etadetom}{\eta_{\mathrm{det}}^{\mathrm{OM}}}

\begin{document}

\title{Observation of Strong Radiation Pressure Forces from Squeezed Light on a Mechanical Oscillator}

\author{Jeremy B. Clark, Florent Lecocq, Raymond W. Simmonds, Jos\'{e} Aumentado, John D. Teufel}
\affiliation{National Institute of Standards and Technology, Boulder, CO 80305 USA}



\begin{abstract}
\end{abstract}

\maketitle

\textbf{Quantum enhanced sensing is a powerful technique in which nonclassical states are used to improve the sensitivity of a measurement \cite{giovannetti_quantum-enhanced_2004}.
	For enhanced mechanical displacement sensing, squeezed states of light have been shown to reduce the “photon counting noise” that limits the measurement noise floor \cite{mckenzie_experimental_2002, vahlbruch_demonstration_2005, goda_quantum-enhanced_2008, aasi_enhanced_2013, taylor_biological_2013, pooser_ultrasensitive_2015}.
	It has long been predicted, however, that suppressing the noise floor with squeezed light should produce an unavoidable increase in radiation pressure noise that drives the mechanical system \cite{caves_quantum-mechanical_1981}.
	Such nonclassical radiation pressure forces have thus far been hidden by insufficient measurement strengths and residual thermal mechanical motion.
	Since the ultimate measurement sensitivity relies on the delicate balance between these two noise sources, the limits of the quantum enhancement have not been observed.
	Using a microwave cavity optomechanical system, we observe the nonclassical radiation pressure noise that necessarily accompanies any quantum enhancement of the measurement precision.
	By varying both the magnitude and phase of the squeezing, we optimize the fundamental trade-off between mechanical imprecision and backaction noise in accordance with the Heisenberg uncertainty principle.
	As the strength of the measurement is further increased, radiation pressure forces eventually dominate the mechanical motion.
	In this regime, the optomechanical interaction can be exploited as an efficient quantum nondemolition (QND) measurement of the amplitude fluctuations of the light field \cite{jacobs_quantum-nondemolition_1994}.
	By overwhelming mechanical thermal noise with radiation pressure by two orders of magnitude, we demonstrate a mechanically-mediated measurement of the squeezing with an effective homodyne efficiency of 94\%.
	Thus, with strong radiation pressures forces, mechanical motion enhances the measurement of nonclassical light, just as nonclassical light enhances the measurement of the motion.
	Looking forward, radiation pressure forces from squeezed light are expected to play an important role in determining the ultimate precision of future gravitational wave observatories \cite{schnabel_quantum_2010}.}

In recent years, precision measurements of mechanical motion have become increasingly limited by quantum effects \cite{braginsky_quantum_1995,clerk_introduction_2010}.
Some prime laboratory examples have come from cavity optomechanical systems wherein the motion of a mechanical oscillator couples to a mode of an electromagnetic cavity resonator \cite{aspelmeyer_cavity_2014, purdy_observation_2013, schreppler_optically_2014, wilson_measurement-based_2015, teufel_overwhelming_2015}.
When the cavity is probed at its resonance frequency with a coherent state, any mechanical motion is transduced to a phase modulation of the returned probe.
The probe's phase noise therefore establishes the detection's noise floor and ultimately limits the measurement's signal-to-noise ratio.
Although this noise floor can be reduced by increasing the drive power, limited power availability or low material damage thresholds occasionally call for the use of squeezed states to push the drive's phase fluctuations below the shot noise limit (SNL).
To date, however, the impact of radiation pressure forces that accompany this reduction in phase fluctuations below the SNL has not been observed.
We investigate this impact by interrogating a cavity optomechanical system using squeezed microwave radiation across regimes of weak and strong radiation pressure forces, characterizing the total measurement noise performance throughout.

\begin{figure*}
	\includegraphics[width = 15cm]{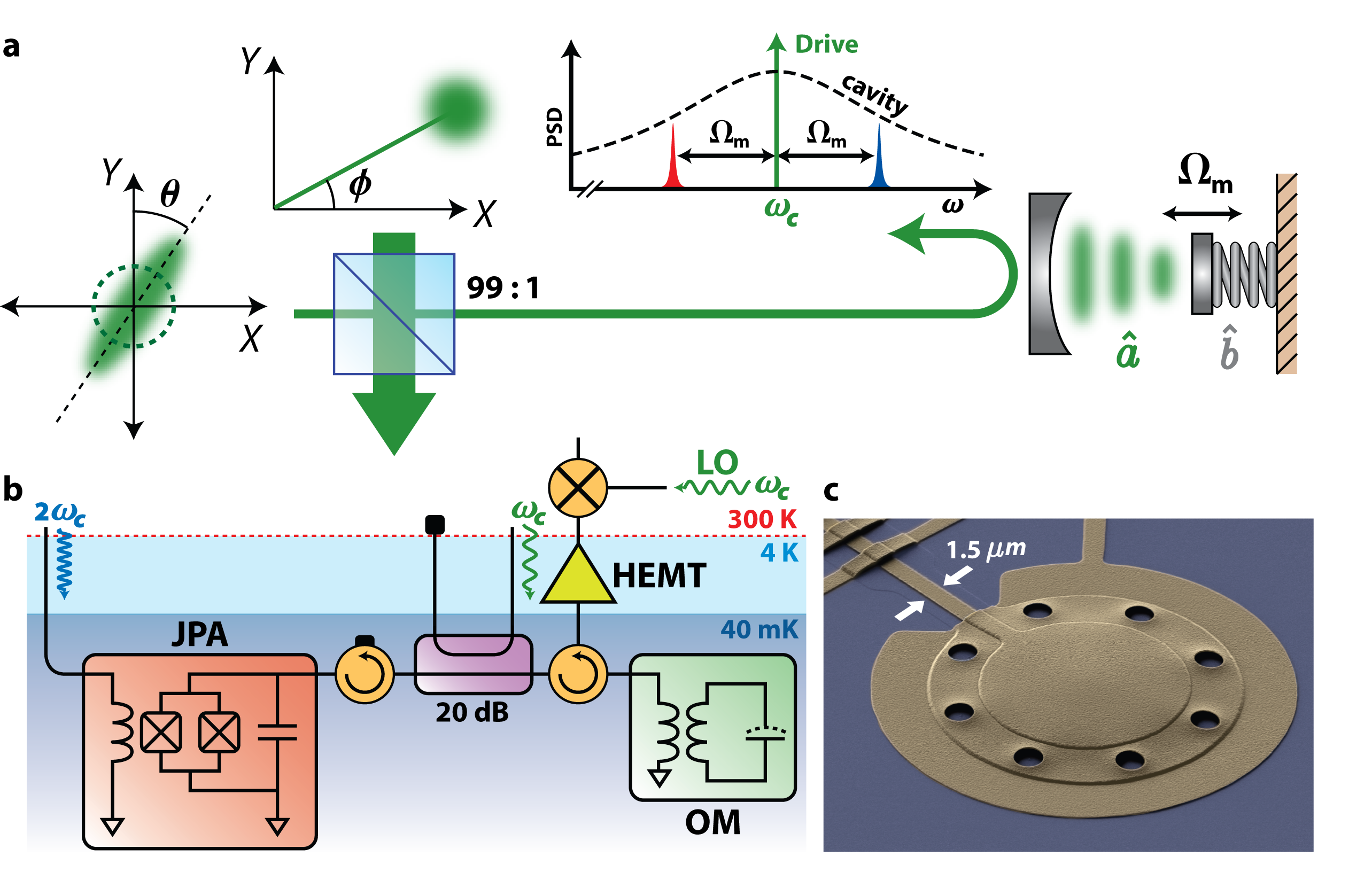}
	\caption{\label{fig:concept}
		Experimental concept.
		\textbf{a}, Optical representation of the experiment.
		A displaced squeezed state is prepared by combining a strong coherent state with a state of squeezed vacuum using a 99:1 beamsplitter.
		The drive field then interrogates the motion of the mechanical oscillator, inducing phase modulation sidebands above and below the reflected drive's carrier frequency.
		\textbf{b}, Equivalent microwave circuit.
		Squeezed vacuum is prepared by parametrically pumping a Josephson Parametric Amplifier (JPA) at twice the resonance frequency of the optomechanical cavity ($2\omega_c$).
		The squeezed state is then combined with the coherent drive via the coupled port of 20~dB directional coupler.
		The optomechanical circuit (OM) is interrogated via a reflection measurement, and the reflected field is then amplified using a phase-insensitive HEMT amplifier.
		\textbf{c}, False-color image of the aluminum drum deposited on a sapphire substrate (blue) taken with a scanning electron microscope.}
\end{figure*}

In our experiments (Fig.~\ref{fig:concept}), the microwave ``cavity" consists of an aluminum 15~nH spiral inductor shunted by a mechanically--compliant vacuum--gap capacitor \cite{cicak_low-loss_2010, teufel_circuit_2011}.
We engineer the total cavity linewidth ($\kappa/2\pi= 22.2~\mathrm{MHz}$) to be on the order of the mechanical resonance frequency of the capacitor's primary flexural mode ($\Omega_{\mathrm{m}}/2\pi=8.68$~MHz) in order to maximize the measurement strength per cavity photon.
The circuit is cooled in a dilution refrigerator and held at a temperature of $T=40$~mK, corresponding to an equilibrium phonon occupancy of approximately 95 quanta.
We further reduce this phonon occupancy by sideband cooling \cite{teufel_sideband_2011} the mechanical oscillator to an occupancy of $\nth=10$ and a mechanical linewidth of $\Gamma/2\pi=200$~Hz throughout all of our experiments (see supplementary information).
A state of microwave squeezed vacuum is prepared \cite{castellanos-beltran_amplification_2008} using a Josephson Parametric Amplifier (JPA) and combined with a strong microwave drive that is tuned to the cavity's resonance frequency ($\omega_c/2\pi=6.89$ GHz).
By manipulating the phase of the squeezing, $\theta$, relative to the phase of the drive, $\phi$, we are able to prepare a variety of displaced squeezed states that we use to interrogate the optomechanical cavity via reflection.
The reflected field is then amplified using a cryogenic high electron mobility transistor (HEMT) amplifier before being demodulated via homodyne detection.

\begin{figure*}
	\includegraphics[width = 15cm]{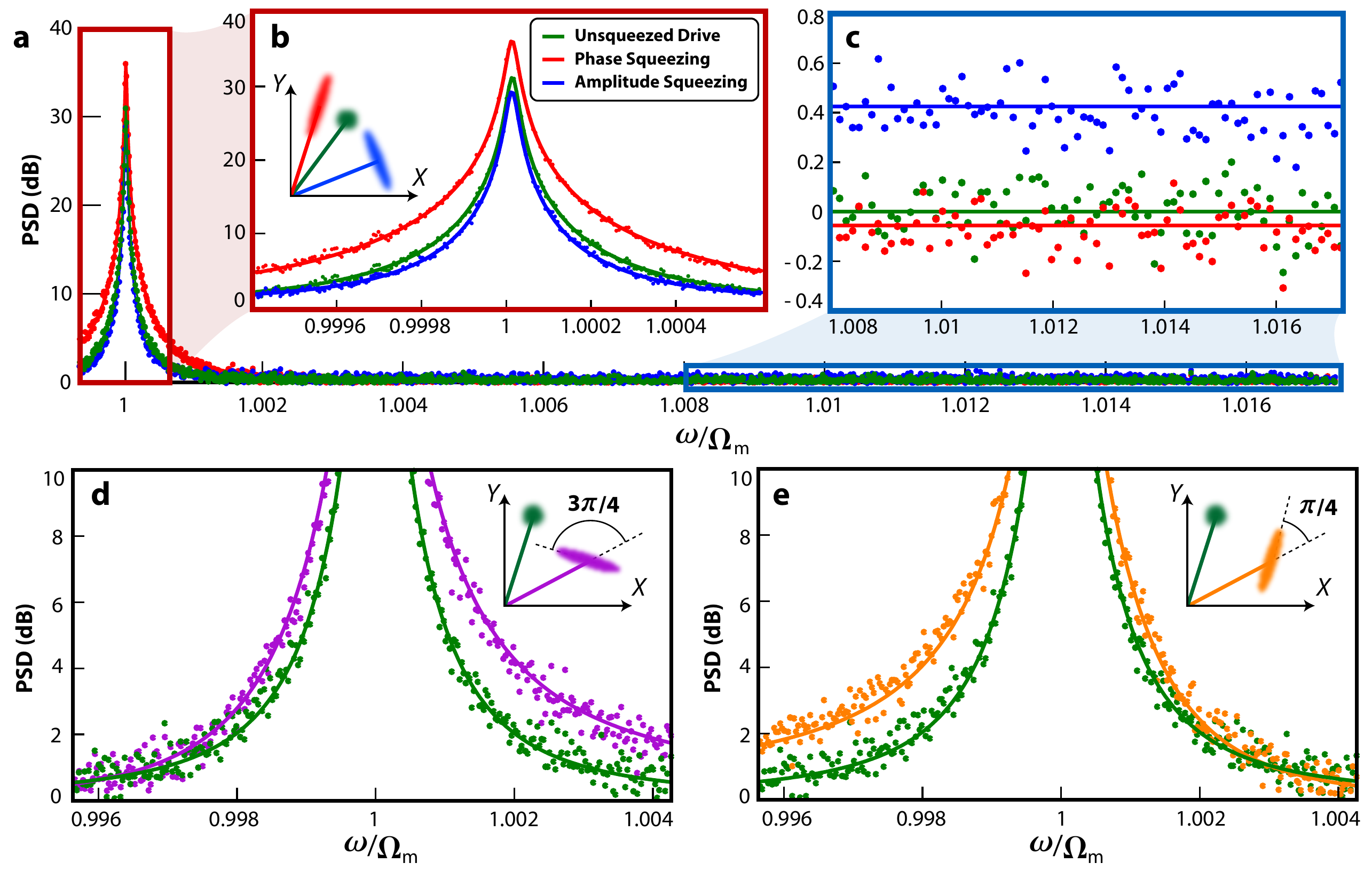}
	\caption{\label{fig:lorentzians}
		Homodyne power spectral density of the detected mechanical sideband for various squeezing angles, $\theta$.
		All plotted spectra are normalized to amplified shot noise.
		\textbf{a}, Broadband mechanical noise spectra in the presence of an unsqueezed drive as well as amplitude and phase squeezing.
		For this data, $C=70$ and $r=0.9$.
		\textbf{b}, Close-up of the Lorentzian mechanical sideband in the presence of various squeezed states.
		The area of each is proportional to the equilibrium mechanical mode temperature.
		Phase squeezing increases the measurement backaction compared to the unsqueezed case, and amplitude squeezing reduces the backaction.
		\textbf{c}, Zoomed-in depiction of the baseband noise away from the mechanical Lorentzian.
		The data have been smoothed using a moving average with a bin width of 30 points.
		The phase-squeezed drive lowers the noise floor while amplitude squeezing raises it.
		\textbf{d}, \textbf{e}, Asymmetric shape of the mechanical resonance at intermediate squeezing phases ($C=220$).
		The plotted theory assumes no free parameters.}
\end{figure*}

Figure~\ref{fig:lorentzians} displays the power spectral density of the demodulated mechanical sidebands in the regime of strong radiation pressure noise.
In this regime, the effects of the squeezing appear in two distinct ways in the spectra.
The peak height of the Lorentzian lineshape at the mechanical resonance frequency, $\Omega_\mathrm{m}$, is largely determined by the magnitude of the amplitude fluctuations of the drive field, which is responsible for the backaction.
In contrast, the measurement noise floor is set by the fluctuations of the phase quadrature of the drive.
At intermediate phases between amplitude and phase squeezing, the correlations of the drive field lead to constructive and destructive interference in the wings of the mechanical line (Fig.~\ref{fig:lorentzians}d and e), yielding a Fano-like lineshape.
It is precisely these correlations that advanced gravitational wave observatories will exploit to enhance the broadband strain sensitivities of their detectors away from the mechanical resonance frequency once radiation pressure forces become important \cite{kimble_conversion_2001, schnabel_quantum_2010, demkowicz_dobrzanski_fundamental_2013}.

The full role of the squeezing can be rigorously understood by considering the interaction Hamiltonian that couples a cavity resonator mode $\hat{a}$ to a mechanical oscillator mode $\hat{b}$:
\begin{equation}
\label{eq:interaction}
\hat{H}_{\mathrm{int}} = \hbar g_0\hat{a}^\dagger\hat{a}(\hat{b}^\dagger+\hat{b}),
\end{equation}
where $g_0$ denotes the vacuum optomechanical coupling rate.
In a linearized regime, a large coherent build-up of the cavity photons $\braket{\hat{a}^\dagger\hat{a}}=|\alpha|^2$ achieves a parametric enhancement of the coupling rate $g=g_0\alpha$, which allows the interaction strength between the light field and the mechanical oscillator to be tuned by changing the drive power.
We will parameterize the coupling rate in terms of the optomechanical cooperativity $C\equiv4g^2/\kappa\Gamma$, where $\Gamma$ and $\kappa$ are the mechanical and cavity linewidths, respectively \cite{aspelmeyer_cavity_2014}.
It is the cooperativity that weights how the fluctuations of the amplitude $\braket{(\Delta\hat{X}_a}^2)$ and phase $\braket{(\Delta\hat{Y}_a)^2}$ quadratures of the drive field contribute to the detection's total added noise.

To characterize the measurement's noise performance, it will be useful to quantify the imprecision noise floor in units of equivalent thermal phonons \cite{wilson_measurement-based_2015}
\begin{equation}
\label{eq:nimp}
\nimp=\frac{1-\eta_{\mathrm{det}}+4\eta_{\mathrm{det}}\braket{(\Delta\hat{Y}_a)^2}}{4\eta_{\mathrm{det}} \widetilde{C}},
\end{equation}
where we have assumed a homodyne detection of quantum efficiency $\eta_{\mathrm{det}}$ \cite{leonhart_2005} and introduced the weighted cooperativity $\widetilde{C}\equiv 4C/(1+4(\tfrac{\Omega_m}{\kappa})^2)$.
As Eq.~\ref{eq:nimp} confirms, $\nimp$ can be suppressed by squeezing the fluctuations of the drive's phase quadrature or by increasing the cooperativity.
The contribution to the thermal phonon occupancy due to measurement backaction, $n_{\mathrm{ba}}$, is proportional to the variance of the light field's amplitude quadrature according to
\begin{equation}
\label{eq:nba}
\nba=\tilde{C}\braket{(\Delta\hat{X}_a)^2}.
\end{equation}
Thus, $\nimp$ and $\nba$ are related to one another by the light field's Heisenberg uncertainty principle  $\braket{(\Delta\hat{X}_a)^2}\braket{(\Delta\hat{Y}_a)^2}\ge1/16$.
When interrogating the cavity with a coherent state, the variance of the drive's amplitude and phase quadratures both obey $\braket{(\Delta\hat{Y}_a)^2}=\braket{(\Delta\hat{X}_a)^2} = \tfrac{1}{4}$, which determines the SNL of the light.
In this case, the noise added by the detection ($\nadd=\nimp + \nba$) is governed solely by the choice of measurement cooperativity, and the minimum value of added measurement noise ($\nadd>\tfrac{1}{2}$) directly follows from the Heisenberg uncertainty principle \cite{caves_measurement_1980, braginsky_quantum_1995, clerk_introduction_2010,aspelmeyer_cavity_2014}.
The introduction of squeezed light provides an additional parameter through which to reach the ultimate limit \cite{arcizet_beating_2006, kerdoncuff_squeezing_enhanced_2015} where $\nadd=1/2$.
As is customary, we will parameterize the squeezing by associating the variance of the drive's squeezed quadrature, $\hat{X}^{\mathrm{sqz}}$, with the squeezing parameter $r$ according to $\braket{(\Delta \hat{X}^{\mathrm{sqz}})^2}/\mathrm{SNL}=e^{-2r}$.
As Eqs.~\ref{eq:nimp} and \ref{eq:nba} illustrate, the imprecision-backaction inequality $\nimp\times\nba\ge1/16$ is maintained in the presence of the squeezing.

\begin{figure*}
	\includegraphics[width = 15cm]{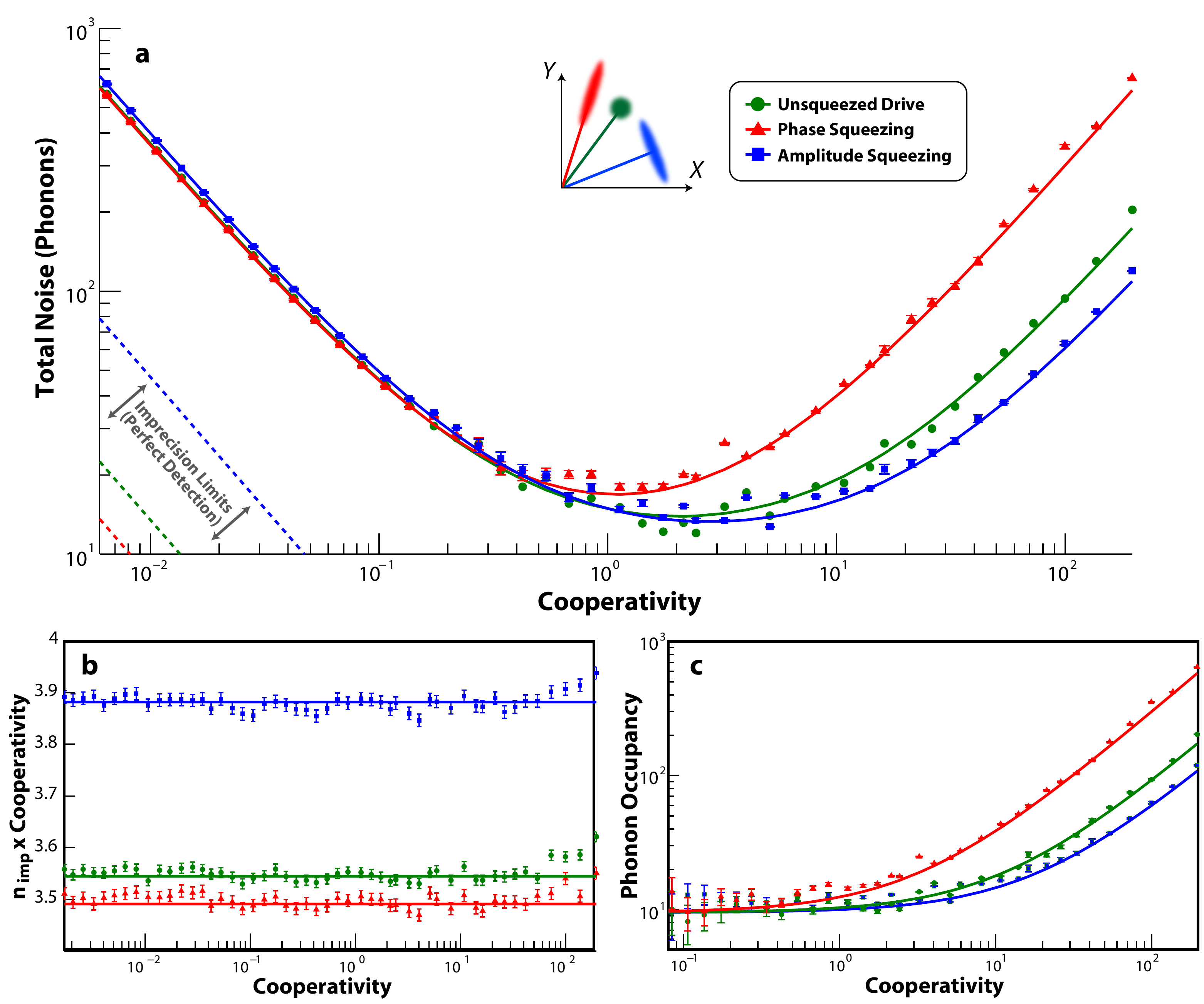}
	\caption{\label{fig:imp_ba}
		Imprecision and backaction noise in the presence of amplitude and phase squeezing ($r=1$).
		The uncertainties are estimated using 68\% confidence intervals for the Lorentzian fit parameters.
		\textbf{a}, Total detected noise, expressed in units of mechanical quanta, versus the measurement cooperativity.
		In the presence of amplitude squeezing (blue squares), we observe more added noise in the imprecision limit and less added noise in the back-action limit compared to an unsqueezed drive (green circles).
		These behaviors are reversed in the presence of a phase-squeezed drive (red triangles).
		The ideal imprecision noise limits that would be obtained with perfect detection efficiency are portrayed in the lower left-hand corner.
		\textbf{b}, Contribution of imprecision noise to the total added noise.
		As the imprecision noise should scale inversely with $C$, we plot the product $\nimp\times C$, which should be constant.
		\textbf{c},  Measured equilibrium phonon occupancy in the presence of the various drive states.}
\end{figure*}

By knowing the vacuum coupling rate ($g_0/2\pi = 170$~Hz) and the full quantum state of the cavity field, it is straightforward to report the measurement's total noise in phonon units ($\nimp + \nba +\nth + 1/2$), which is portrayed Fig.~\ref{fig:imp_ba}.
In particular, we must consider the effects of a limited microwave transmittance between the JPA and the optomechanical circuit ($\etain=47\pm2\%$) and a constant thermal occupancy of the cavity field ($n_c=0.17$), which act to limit the purity of the squeezing (see supplementary material).
The data show full quantitative agreement with the expected noise performance in all measurement regimes, including those where either $\nimp$ or $\nba$ dominate the detection's total noise.
Notably, we observe a nearly 3~dB reduction of the total measurement noise by using amplitude squeezing in the backaction limit ($C\gg\nth$).
In the case where $\nimp=\nba$, the purity of the drive field becomes important in achieving a minimum level of added noise.
In principle, a sufficiently pure amplitude-squeezed state mitigates the effects of a finite detection efficiency \cite{kerdoncuff_squeezing_enhanced_2015}, which allows the measurement to more closely approach the ultimate quantum limit where $\nadd=1/2$.

Just as the squeezing can be used to enhance the detection of the mechanical oscillator, the mechanical oscillator can be used to enhance the detection of the squeezing.
In either case, the characteristics of the measurement are embodied by the interaction Hamiltonian (Eq.~\ref{eq:interaction}).
Here, the coupled mechanical observable $\hat{X}_b=\hat{b}^\dagger+\hat{b}$ fails to commute with its bare Hamiltonian $\hat{b}^\dagger\hat{b}$, causing any measurement disturbance to feed back into $\hat{X}_b$ during the course of the system's evolution.
This should be contrasted with the coupled cavity field observable $\hat{a}^\dagger\hat{a}$, which suggests that the mechanical oscillator performs a QND measurement \cite{braginsky_quantum_1980} of the cavity photon number.
In the linearized regime, we take $\hat{a} \rightarrow \alpha +\delta\hat{a}$, which transforms the QND observable to the fluctuations of the amplitude quadrature, $\delta\hat{X}_a$, since $\hat{a}^\dagger\hat{a}\rightarrow\alpha(\delta\hat{a}^\dagger+\delta\hat{a})$.
It is in this sense that the phonon field state performs a nondestructive detection of the light's amplitude quadrature \cite{jacobs_quantum-nondemolition_1994}.

Typically, a light field's quantum state is tomographically reconstructed by detecting an ensemble of identically prepared states with a homodyne receiver.
By rotating the phase of the receiver's local oscillator, various quadratures of the light field can be selected over the course of many experiments, and the quantum state of the field can be reconstructed from the statistics.
Here, since the mechanical mode is only sensitive to the drive's amplitude fluctuations, we rotate the phase of the squeezing, $\theta$, relative to the phase of the coherent drive, $\phi$, to achieve the same goal.
The equilibrium mechanical occupancy is sensitive to both the magnitude and phase of the squeezing, so each mechanical sideband contains information about both $r$ and $\theta$.
Thus, we are able to completely characterize the squeezed state using a heterodyne measurement of the returned drive's upper mechanical sideband.
Unlike conventional homodyne detection, however, the QND measurement pursued here does not destroy the light field upon its detection and ideally leaves the amplitude quadrature unperturbed by any measurement backaction (see supplementary information).

\begin{figure*}
	\includegraphics[width = 15cm]{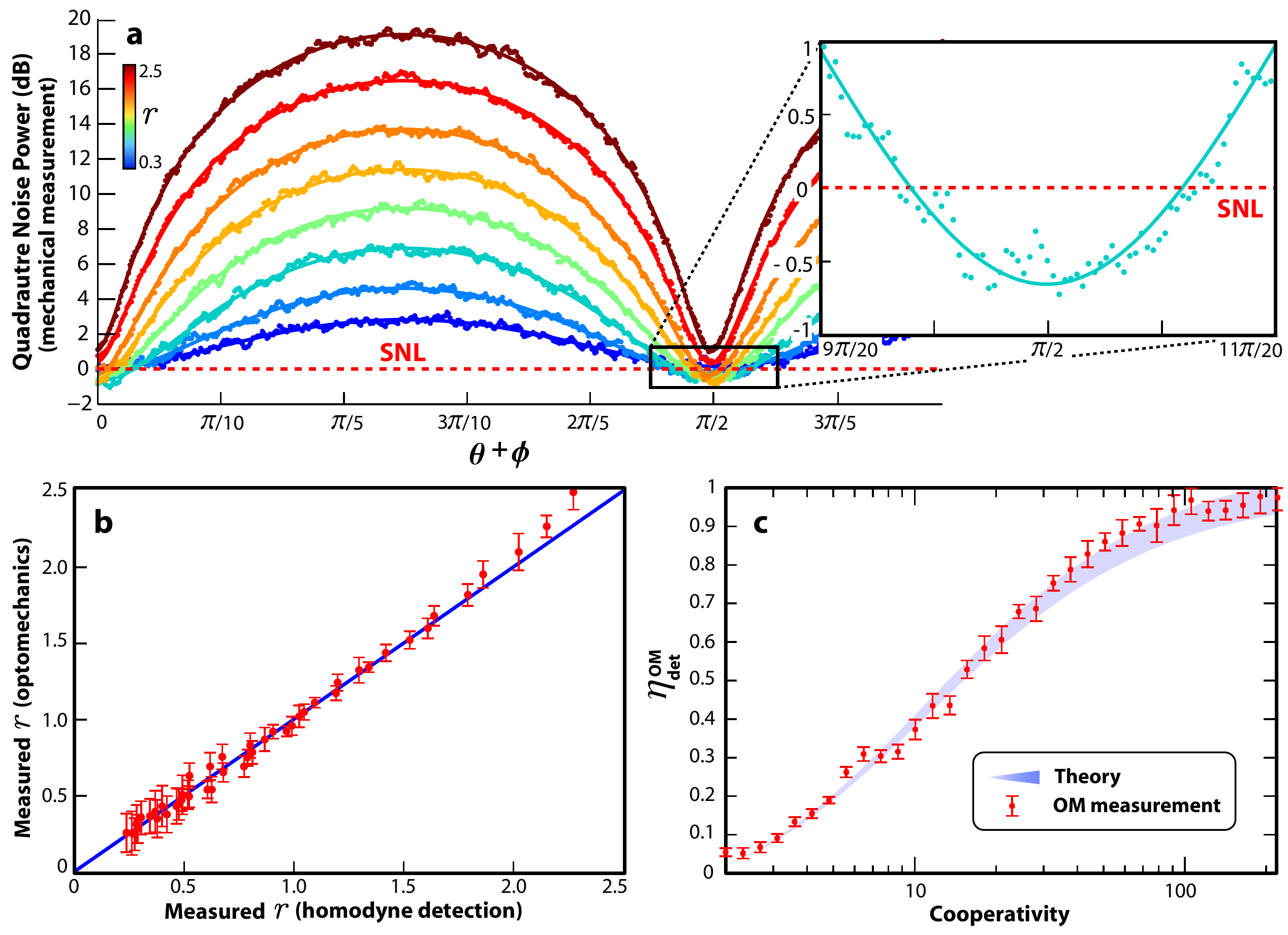}
	\caption{\label{fig:detector}
		Detecting microwave squeezing using the mechanical state.
		\textbf{a}, Average integrated noise power of the upper mechanical sideband (measured via heterodyne detection) as a function of the squeezing phase.
		The colors denote different JPA drive powers and, correspondingly, different values of the squeezing parameter, $r$.
		The fits assume a squeezed thermal state subject to loss (see supplementary information).
		The squeezing shown in the inset is comparable to that used to obtain the data in Fig.~\ref{fig:imp_ba}.
		\textbf{b}, Squeezing parameter, $r$, measured using the optomechanical cavity versus similar results obtained using conventional homodyne detection performed with a large integration bandwidth (1000~$\times~\Gamma$).
		The blue line depicts a slope of unity.
		\textbf{c}, Effective mechanical detection efficiency of the squeezing, $\etadet^{\mathrm{OM}}$, as a function of drive cooperativity (see Eq.~\ref{eq:etadetom}).
		The expected range of $\etadet^{\mathrm{OM}}$ values (shaded blue) accounts for the uncertainty in the microwave transmittance ($\etain=47\pm2\%$) between the Josephson parametric amplifier and the optomechanical cavity.
		The uncertainties in (b) and (c) correspond to the standard deviation of the mean over 20 experiments.}
\end{figure*}

For the mechanical mode to facilitate an efficient detection of the squeezing, it is critical that the rate at which pump photons are scattered into the mechanical sidebands, $\Gamma_{\mathrm{scatter}}=4g^2\kappa/(\kappa^2+4\Omega_m^2)$, greatly exceed the rate of mechanical decoherence, $\nth\Gamma$.
Under these conditions, the state of the mechanical oscillator couples more strongly to the quantum correlations of the squeezed microwave field than the mechanical thermal environment.
To demonstrate this idea, we measure the total noise power of the upper mechanical sideband at various measurement cooperativities as the squeezing phase, $\theta$, is rotated.
By normalizing the phase--dependent mechanical noise power to the detected level of mechanical noise in the presence of a coherent state drive, we reconstruct the squeezed state's quadrature noise power as in Fig.~\ref{fig:detector}a.
Just as with conventional homodyne detection, the measured level of squeezing can be quantified by an effective detection efficiency, $\etadet^{\mathrm{OM}}$, according to
\begin{equation}
\label{eq:etadetom}
\frac{\braket{(\Delta\hat{X}_a)^2}_{\mathrm{measured}}}{\mathrm{SNL}}=1-\etaeff^{\mathrm{OM}}+\etaeff^{\mathrm{OM}}\times\frac{\braket{(\Delta\hat{X}_a)^2}_{\mathrm{input}}}{\mathrm{SNL}}.
\end{equation}
At high drive strengths where $\nth\gg\nimp$, $\etaeff^{\mathrm{OM}}$ scales with cooperativity according to $\etadet^{\mathrm{OM}}=(1+\nth\Gamma/\Gamma_{\mathrm{scatter}})^{-1}$.
At the highest measurement cooperativities accessible in our experiments, we expect $\etadet^{\mathrm{OM}}\approx94\%$, which is confirmed in Fig.~\ref{fig:detector}b.
This amounts to a 30-fold improvement over the homodyne efficiency of the conventional microwave detection chain used in these experiments ($\eta_{\mathrm{det}}=3\%$).

These measurements demonstrate both how quantum states of the light field can improve mechanical measurements and how mechanical systems can make improved measurements of the light.
Looking forward, squeezed light could prove to be a valuable resource for the preparation of squeezed or entangled \cite{huang_entangling_2009} states of motion.
Furthermore, cavity optomechanical systems in the backaction-dominated regime offer an ideal testbed for quantum enhanced displacement sensing with other nonclassical states of light \cite{giovannetti_quantum-enhanced_2004, demkowicz_dobrzanski_fundamental_2013}.
Lastly, the QND nature of the measurement of the light field also opens the door for novel types of mechanically-mediated noiseless amplification \cite{metelmann_quantum-limited_2014}.



%



\begin{acknowledgments}
This work was supported by NIST and the DARPA QuASAR program.  We thank M. A. Castellanos-Beltran and A. J. Sirois for valuable conversations and technical assistance,  L. R. Vale for fabrication of the JPA, and
A.W. Sanders for taking the SEM micrograph in Fig.~\ref{fig:concept}c.  J.B. Clark acknowledges the NRC for financial support.
\end{acknowledgments}

\bibliography{sqz_int_bib}

\begin{thebibliography}{30}%
\makeatletter
\providecommand \@ifxundefined [1]{%
 \@ifx{#1\undefined}
}%
\providecommand \@ifnum [1]{%
 \ifnum #1\expandafter \@firstoftwo
 \else \expandafter \@secondoftwo
 \fi
}%
\providecommand \@ifx [1]{%
 \ifx #1\expandafter \@firstoftwo
 \else \expandafter \@secondoftwo
 \fi
}%
\providecommand \natexlab [1]{#1}%
\providecommand \enquote  [1]{``#1''}%
\providecommand \bibnamefont  [1]{#1}%
\providecommand \bibfnamefont [1]{#1}%
\providecommand \citenamefont [1]{#1}%
\providecommand \href@noop [0]{\@secondoftwo}%
\providecommand \href [0]{\begingroup \@sanitize@url \@href}%
\providecommand \@href[1]{\@@startlink{#1}\@@href}%
\providecommand \@@href[1]{\endgroup#1\@@endlink}%
\providecommand \@sanitize@url [0]{\catcode `\\12\catcode `\$12\catcode
  `\&12\catcode `\#12\catcode `\^12\catcode `\_12\catcode `\%12\relax}%
\providecommand \@@startlink[1]{}%
\providecommand \@@endlink[0]{}%
\providecommand \url  [0]{\begingroup\@sanitize@url \@url }%
\providecommand \@url [1]{\endgroup\@href {#1}{\urlprefix }}%
\providecommand \urlprefix  [0]{URL }%
\providecommand \Eprint [0]{\href }%
\providecommand \doibase [0]{http://dx.doi.org/}%
\providecommand \selectlanguage [0]{\@gobble}%
\providecommand \bibinfo  [0]{\@secondoftwo}%
\providecommand \bibfield  [0]{\@secondoftwo}%
\providecommand \translation [1]{[#1]}%
\providecommand \BibitemOpen [0]{}%
\providecommand \bibitemStop [0]{}%
\providecommand \bibitemNoStop [0]{.\EOS\space}%
\providecommand \EOS [0]{\spacefactor3000\relax}%
\providecommand \BibitemShut  [1]{\csname bibitem#1\endcsname}%
\let\auto@bib@innerbib\@empty
\bibitem [{\citenamefont {Giovannetti}\ \emph {et~al.}(2004)\citenamefont
  {Giovannetti}, \citenamefont {Lloyd},\ and\ \citenamefont
  {Maccone}}]{giovannetti_quantum-enhanced_2004}%
  \BibitemOpen
  \bibfield  {author} {\bibinfo {author} {\bibfnamefont {V.}~\bibnamefont
  {Giovannetti}}, \bibinfo {author} {\bibfnamefont {S.}~\bibnamefont {Lloyd}},
  \ and\ \bibinfo {author} {\bibfnamefont {L.}~\bibnamefont {Maccone}},\
  }\href@noop {} {\bibfield  {journal} {\bibinfo  {journal} {Science}\ }\textbf
  {\bibinfo {volume} {306}},\ \bibinfo {pages} {1330} (\bibinfo {year}
  {2004})}\BibitemShut {NoStop}%
\bibitem [{\citenamefont {McKenzie}\ \emph {et~al.}(2002)\citenamefont
  {McKenzie}, \citenamefont {Shaddock}, \citenamefont {McClelland},
  \citenamefont {Buchler},\ and\ \citenamefont
  {Lam}}]{mckenzie_experimental_2002}%
  \BibitemOpen
  \bibfield  {author} {\bibinfo {author} {\bibfnamefont {K.}~\bibnamefont
  {McKenzie}}, \bibinfo {author} {\bibfnamefont {D.~A.}\ \bibnamefont
  {Shaddock}}, \bibinfo {author} {\bibfnamefont {D.~E.}\ \bibnamefont
  {McClelland}}, \bibinfo {author} {\bibfnamefont {B.~C.}\ \bibnamefont
  {Buchler}}, \ and\ \bibinfo {author} {\bibfnamefont {P.~K.}\ \bibnamefont
  {Lam}},\ }\href@noop {} {\bibfield  {journal} {\bibinfo  {journal} {Phys.
  Rev. Lett.}\ }\textbf {\bibinfo {volume} {88}},\ \bibinfo {pages} {231102}
  (\bibinfo {year} {2002})}\BibitemShut {NoStop}%
\bibitem [{\citenamefont {Vahlbruch}\ \emph {et~al.}(2005)\citenamefont
  {Vahlbruch}, \citenamefont {Chelkowski}, \citenamefont {Hage}, \citenamefont
  {Franzen}, \citenamefont {Danzmann},\ and\ \citenamefont
  {Schnabel}}]{vahlbruch_demonstration_2005}%
  \BibitemOpen
  \bibfield  {author} {\bibinfo {author} {\bibfnamefont {H.}~\bibnamefont
  {Vahlbruch}}, \bibinfo {author} {\bibfnamefont {S.}~\bibnamefont
  {Chelkowski}}, \bibinfo {author} {\bibfnamefont {B.}~\bibnamefont {Hage}},
  \bibinfo {author} {\bibfnamefont {A.}~\bibnamefont {Franzen}}, \bibinfo
  {author} {\bibfnamefont {K.}~\bibnamefont {Danzmann}}, \ and\ \bibinfo
  {author} {\bibfnamefont {R.}~\bibnamefont {Schnabel}},\ }\href@noop {}
  {\bibfield  {journal} {\bibinfo  {journal} {Phys. Rev. Lett.}\ }\textbf
  {\bibinfo {volume} {95}},\ \bibinfo {pages} {211102} (\bibinfo {year}
  {2005})}\BibitemShut {NoStop}%
\bibitem [{\citenamefont {Goda}\ \emph {et~al.}(2008)\citenamefont {Goda},
  \citenamefont {Miyakawa}, \citenamefont {Mikhailov}, \citenamefont {Saraf},
  \citenamefont {Adhikari}, \citenamefont {McKenzie}, \citenamefont {Ward},
  \citenamefont {Vass}, \citenamefont {Weinstein},\ and\ \citenamefont
  {Mavalvala}}]{goda_quantum-enhanced_2008}%
  \BibitemOpen
  \bibfield  {author} {\bibinfo {author} {\bibfnamefont {K.}~\bibnamefont
  {Goda}}, \bibinfo {author} {\bibfnamefont {O.}~\bibnamefont {Miyakawa}},
  \bibinfo {author} {\bibfnamefont {E.~E.}\ \bibnamefont {Mikhailov}}, \bibinfo
  {author} {\bibfnamefont {S.}~\bibnamefont {Saraf}}, \bibinfo {author}
  {\bibfnamefont {R.}~\bibnamefont {Adhikari}}, \bibinfo {author}
  {\bibfnamefont {K.}~\bibnamefont {McKenzie}}, \bibinfo {author}
  {\bibfnamefont {R.}~\bibnamefont {Ward}}, \bibinfo {author} {\bibfnamefont
  {S.}~\bibnamefont {Vass}}, \bibinfo {author} {\bibfnamefont {A.~J.}\
  \bibnamefont {Weinstein}}, \ and\ \bibinfo {author} {\bibfnamefont
  {N.}~\bibnamefont {Mavalvala}},\ }\href@noop {} {\bibfield  {journal}
  {\bibinfo  {journal} {Nat Phys}\ }\textbf {\bibinfo {volume} {4}},\ \bibinfo
  {pages} {472} (\bibinfo {year} {2008})}\BibitemShut {NoStop}%
\bibitem [{\citenamefont {{The LIGO Scientific
  Collaboration}}(2013)}]{aasi_enhanced_2013}%
  \BibitemOpen
  \bibfield  {author} {\bibinfo {author} {\bibnamefont {{The LIGO Scientific
  Collaboration}}},\ }\href@noop {} {\bibfield  {journal} {\bibinfo  {journal}
  {Nat Photon}\ }\textbf {\bibinfo {volume} {7}},\ \bibinfo {pages} {613}
  (\bibinfo {year} {2013})}\BibitemShut {NoStop}%
\bibitem [{\citenamefont {Taylor}\ \emph {et~al.}(2013)\citenamefont {Taylor},
  \citenamefont {Janousek}, \citenamefont {Daria}, \citenamefont {Knittel},
  \citenamefont {Hage}, \citenamefont {Bachor},\ and\ \citenamefont
  {Bowen}}]{taylor_biological_2013}%
  \BibitemOpen
  \bibfield  {author} {\bibinfo {author} {\bibfnamefont {M.~A.}\ \bibnamefont
  {Taylor}}, \bibinfo {author} {\bibfnamefont {J.}~\bibnamefont {Janousek}},
  \bibinfo {author} {\bibfnamefont {V.}~\bibnamefont {Daria}}, \bibinfo
  {author} {\bibfnamefont {J.}~\bibnamefont {Knittel}}, \bibinfo {author}
  {\bibfnamefont {B.}~\bibnamefont {Hage}}, \bibinfo {author} {\bibfnamefont
  {H.-A.}\ \bibnamefont {Bachor}}, \ and\ \bibinfo {author} {\bibfnamefont
  {W.~P.}\ \bibnamefont {Bowen}},\ }\href@noop {} {\bibfield  {journal}
  {\bibinfo  {journal} {Nat Photon}\ }\textbf {\bibinfo {volume} {7}},\
  \bibinfo {pages} {229} (\bibinfo {year} {2013})}\BibitemShut {NoStop}%
\bibitem [{\citenamefont {Pooser}\ and\ \citenamefont
  {Lawrie}(2015)}]{pooser_ultrasensitive_2015}%
  \BibitemOpen
  \bibfield  {author} {\bibinfo {author} {\bibfnamefont {R.~C.}\ \bibnamefont
  {Pooser}}\ and\ \bibinfo {author} {\bibfnamefont {B.}~\bibnamefont
  {Lawrie}},\ }\href@noop {} {\bibfield  {journal} {\bibinfo  {journal}
  {Optica, OPTICA}\ }\textbf {\bibinfo {volume} {2}},\ \bibinfo {pages} {393}
  (\bibinfo {year} {2015})}\BibitemShut {NoStop}%
\bibitem [{\citenamefont {Caves}(1981)}]{caves_quantum-mechanical_1981}%
  \BibitemOpen
  \bibfield  {author} {\bibinfo {author} {\bibfnamefont {C.~M.}\ \bibnamefont
  {Caves}},\ }\href@noop {} {\bibfield  {journal} {\bibinfo  {journal} {Phys.
  Rev. D}\ }\textbf {\bibinfo {volume} {23}},\ \bibinfo {pages} {1693}
  (\bibinfo {year} {1981})}\BibitemShut {NoStop}%
\bibitem [{\citenamefont {Jacobs}\ \emph {et~al.}(1994)\citenamefont {Jacobs},
  \citenamefont {Tombesi}, \citenamefont {Collett},\ and\ \citenamefont
  {Walls}}]{jacobs_quantum-nondemolition_1994}%
  \BibitemOpen
  \bibfield  {author} {\bibinfo {author} {\bibfnamefont {K.}~\bibnamefont
  {Jacobs}}, \bibinfo {author} {\bibfnamefont {P.}~\bibnamefont {Tombesi}},
  \bibinfo {author} {\bibfnamefont {M.~J.}\ \bibnamefont {Collett}}, \ and\
  \bibinfo {author} {\bibfnamefont {D.~F.}\ \bibnamefont {Walls}},\ }\href@noop
  {} {\bibfield  {journal} {\bibinfo  {journal} {Phys. Rev. A}\ }\textbf
  {\bibinfo {volume} {49}},\ \bibinfo {pages} {1961} (\bibinfo {year}
  {1994})}\BibitemShut {NoStop}%
\bibitem [{\citenamefont {Schnabel}\ \emph {et~al.}(2010)\citenamefont
  {Schnabel}, \citenamefont {Mavalvala}, \citenamefont {McClelland},\ and\
  \citenamefont {Lam}}]{schnabel_quantum_2010}%
  \BibitemOpen
  \bibfield  {author} {\bibinfo {author} {\bibfnamefont {R.}~\bibnamefont
  {Schnabel}}, \bibinfo {author} {\bibfnamefont {N.}~\bibnamefont {Mavalvala}},
  \bibinfo {author} {\bibfnamefont {D.~E.}\ \bibnamefont {McClelland}}, \ and\
  \bibinfo {author} {\bibfnamefont {P.~K.}\ \bibnamefont {Lam}},\ }\href@noop
  {} {\bibfield  {journal} {\bibinfo  {journal} {Nat Commun}\ }\textbf
  {\bibinfo {volume} {1}},\ \bibinfo {pages} {121} (\bibinfo {year}
  {2010})}\BibitemShut {NoStop}%
\bibitem [{\citenamefont {Braginsky}\ \emph {et~al.}(1995)\citenamefont
  {Braginsky}, \citenamefont {Khalili},\ and\ \citenamefont
  {Thorne}}]{braginsky_quantum_1995}%
  \BibitemOpen
  \bibfield  {author} {\bibinfo {author} {\bibfnamefont {V.~B.}\ \bibnamefont
  {Braginsky}}, \bibinfo {author} {\bibfnamefont {F.~Y.}\ \bibnamefont
  {Khalili}}, \ and\ \bibinfo {author} {\bibfnamefont {K.~S.}\ \bibnamefont
  {Thorne}},\ }\href@noop {} {\emph {\bibinfo {title} {Quantum
  {Measurement}}}},\ \bibinfo {edition} {1st}\ ed.\ (\bibinfo  {publisher}
  {Cambridge University Press},\ \bibinfo {address} {Cambridge},\ \bibinfo
  {year} {1995})\BibitemShut {NoStop}%
\bibitem [{\citenamefont {Clerk}\ \emph {et~al.}(2010)\citenamefont {Clerk},
  \citenamefont {Devoret}, \citenamefont {Girvin}, \citenamefont {Marquardt},\
  and\ \citenamefont {Schoelkopf}}]{clerk_introduction_2010}%
  \BibitemOpen
  \bibfield  {author} {\bibinfo {author} {\bibfnamefont {A.~A.}\ \bibnamefont
  {Clerk}}, \bibinfo {author} {\bibfnamefont {M.~H.}\ \bibnamefont {Devoret}},
  \bibinfo {author} {\bibfnamefont {S.~M.}\ \bibnamefont {Girvin}}, \bibinfo
  {author} {\bibfnamefont {F.}~\bibnamefont {Marquardt}}, \ and\ \bibinfo
  {author} {\bibfnamefont {R.~J.}\ \bibnamefont {Schoelkopf}},\ }\href@noop {}
  {\bibfield  {journal} {\bibinfo  {journal} {Rev. Mod. Phys.}\ }\textbf
  {\bibinfo {volume} {82}},\ \bibinfo {pages} {1155} (\bibinfo {year}
  {2010})}\BibitemShut {NoStop}%
\bibitem [{\citenamefont {Aspelmeyer}\ \emph {et~al.}(2014)\citenamefont
  {Aspelmeyer}, \citenamefont {Kippenberg},\ and\ \citenamefont
  {Marquardt}}]{aspelmeyer_cavity_2014}%
  \BibitemOpen
  \bibfield  {author} {\bibinfo {author} {\bibfnamefont {M.}~\bibnamefont
  {Aspelmeyer}}, \bibinfo {author} {\bibfnamefont {T.~J.}\ \bibnamefont
  {Kippenberg}}, \ and\ \bibinfo {author} {\bibfnamefont {F.}~\bibnamefont
  {Marquardt}},\ }\href@noop {} {\bibfield  {journal} {\bibinfo  {journal}
  {Rev. Mod. Phys.}\ }\textbf {\bibinfo {volume} {86}},\ \bibinfo {pages}
  {1391} (\bibinfo {year} {2014})}\BibitemShut {NoStop}%
\bibitem [{\citenamefont {Purdy}\ \emph {et~al.}(2013)\citenamefont {Purdy},
  \citenamefont {Peterson},\ and\ \citenamefont
  {Regal}}]{purdy_observation_2013}%
  \BibitemOpen
  \bibfield  {author} {\bibinfo {author} {\bibfnamefont {T.~P.}\ \bibnamefont
  {Purdy}}, \bibinfo {author} {\bibfnamefont {R.~W.}\ \bibnamefont {Peterson}},
  \ and\ \bibinfo {author} {\bibfnamefont {C.~A.}\ \bibnamefont {Regal}},\
  }\href@noop {} {\bibfield  {journal} {\bibinfo  {journal} {Science}\ }\textbf
  {\bibinfo {volume} {339}} (\bibinfo {year} {2013})}\BibitemShut {NoStop}%
\bibitem [{\citenamefont {Schreppler}\ \emph {et~al.}(2014)\citenamefont
  {Schreppler}, \citenamefont {Spethmann}, \citenamefont {Brahms},
  \citenamefont {Botter}, \citenamefont {Barrios},\ and\ \citenamefont
  {Stamper-Kurn}}]{schreppler_optically_2014}%
  \BibitemOpen
  \bibfield  {author} {\bibinfo {author} {\bibfnamefont {S.}~\bibnamefont
  {Schreppler}}, \bibinfo {author} {\bibfnamefont {N.}~\bibnamefont
  {Spethmann}}, \bibinfo {author} {\bibfnamefont {N.}~\bibnamefont {Brahms}},
  \bibinfo {author} {\bibfnamefont {T.}~\bibnamefont {Botter}}, \bibinfo
  {author} {\bibfnamefont {M.}~\bibnamefont {Barrios}}, \ and\ \bibinfo
  {author} {\bibfnamefont {D.~M.}\ \bibnamefont {Stamper-Kurn}},\ }\href@noop
  {} {\bibfield  {journal} {\bibinfo  {journal} {Science}\ }\textbf {\bibinfo
  {volume} {344}},\ \bibinfo {pages} {1486} (\bibinfo {year}
  {2014})}\BibitemShut {NoStop}%
\bibitem [{\citenamefont {Wilson}\ \emph {et~al.}(2015)\citenamefont {Wilson},
  \citenamefont {Sudhir}, \citenamefont {Piro}, \citenamefont {Schilling},
  \citenamefont {Ghadimi},\ and\ \citenamefont
  {Kippenberg}}]{wilson_measurement-based_2015}%
  \BibitemOpen
  \bibfield  {author} {\bibinfo {author} {\bibfnamefont {D.~J.}\ \bibnamefont
  {Wilson}}, \bibinfo {author} {\bibfnamefont {V.}~\bibnamefont {Sudhir}},
  \bibinfo {author} {\bibfnamefont {N.}~\bibnamefont {Piro}}, \bibinfo {author}
  {\bibfnamefont {R.}~\bibnamefont {Schilling}}, \bibinfo {author}
  {\bibfnamefont {A.}~\bibnamefont {Ghadimi}}, \ and\ \bibinfo {author}
  {\bibfnamefont {T.~J.}\ \bibnamefont {Kippenberg}},\ }\href@noop {}
  {\bibfield  {journal} {\bibinfo  {journal} {Nature}\ }\textbf {\bibinfo
  {volume} {524}},\ \bibinfo {pages} {325} (\bibinfo {year}
  {2015})}\BibitemShut {NoStop}%
\bibitem [{\citenamefont {Teufel}\ \emph {et~al.}(2016)\citenamefont {Teufel},
  \citenamefont {Lecocq},\ and\ \citenamefont
  {Simmonds}}]{teufel_overwhelming_2015}%
  \BibitemOpen
  \bibfield  {author} {\bibinfo {author} {\bibfnamefont {J.~D.}\ \bibnamefont
  {Teufel}}, \bibinfo {author} {\bibfnamefont {F.}~\bibnamefont {Lecocq}}, \
  and\ \bibinfo {author} {\bibfnamefont {R.~W.}\ \bibnamefont {Simmonds}},\
  }\href@noop {} {\bibfield  {journal} {\bibinfo  {journal} {Phys. Rev. Lett.}\
  }\textbf {\bibinfo {volume} {116}},\ \bibinfo {pages} {013602} (\bibinfo
  {year} {2016})}\BibitemShut {NoStop}%
\bibitem [{\citenamefont {Cicak}\ \emph {et~al.}(2010)\citenamefont {Cicak},
  \citenamefont {Li}, \citenamefont {Strong}, \citenamefont {Allman},
  \citenamefont {Altomare}, \citenamefont {Sirois}, \citenamefont {Whittaker},
  \citenamefont {Teufel},\ and\ \citenamefont
  {Simmonds}}]{cicak_low-loss_2010}%
  \BibitemOpen
  \bibfield  {author} {\bibinfo {author} {\bibfnamefont {K.}~\bibnamefont
  {Cicak}}, \bibinfo {author} {\bibfnamefont {D.}~\bibnamefont {Li}}, \bibinfo
  {author} {\bibfnamefont {J.~A.}\ \bibnamefont {Strong}}, \bibinfo {author}
  {\bibfnamefont {M.~S.}\ \bibnamefont {Allman}}, \bibinfo {author}
  {\bibfnamefont {F.}~\bibnamefont {Altomare}}, \bibinfo {author}
  {\bibfnamefont {A.~J.}\ \bibnamefont {Sirois}}, \bibinfo {author}
  {\bibfnamefont {J.~D.}\ \bibnamefont {Whittaker}}, \bibinfo {author}
  {\bibfnamefont {J.~D.}\ \bibnamefont {Teufel}}, \ and\ \bibinfo {author}
  {\bibfnamefont {R.~W.}\ \bibnamefont {Simmonds}},\ }\href@noop {} {\bibfield
  {journal} {\bibinfo  {journal} {Applied Physics Letters}\ }\textbf {\bibinfo
  {volume} {96}},\ \bibinfo {pages} {093502} (\bibinfo {year}
  {2010})}\BibitemShut {NoStop}%
\bibitem [{\citenamefont {Teufel}\ \emph
  {et~al.}(2011{\natexlab{a}})\citenamefont {Teufel}, \citenamefont {Li},
  \citenamefont {Allman}, \citenamefont {Cicak}, \citenamefont {Sirois},
  \citenamefont {Whittaker},\ and\ \citenamefont
  {Simmonds}}]{teufel_circuit_2011}%
  \BibitemOpen
  \bibfield  {author} {\bibinfo {author} {\bibfnamefont {J.~D.}\ \bibnamefont
  {Teufel}}, \bibinfo {author} {\bibfnamefont {D.}~\bibnamefont {Li}}, \bibinfo
  {author} {\bibfnamefont {M.~S.}\ \bibnamefont {Allman}}, \bibinfo {author}
  {\bibfnamefont {K.}~\bibnamefont {Cicak}}, \bibinfo {author} {\bibfnamefont
  {A.~J.}\ \bibnamefont {Sirois}}, \bibinfo {author} {\bibfnamefont {J.~D.}\
  \bibnamefont {Whittaker}}, \ and\ \bibinfo {author} {\bibfnamefont {R.~W.}\
  \bibnamefont {Simmonds}},\ }\href@noop {} {\bibfield  {journal} {\bibinfo
  {journal} {Nature}\ }\textbf {\bibinfo {volume} {471}},\ \bibinfo {pages}
  {204} (\bibinfo {year} {2011}{\natexlab{a}})}\BibitemShut {NoStop}%
\bibitem [{\citenamefont {Teufel}\ \emph
  {et~al.}(2011{\natexlab{b}})\citenamefont {Teufel}, \citenamefont {Donner},
  \citenamefont {Li}, \citenamefont {Harlow}, \citenamefont {Allman},
  \citenamefont {Cicak}, \citenamefont {Sirois}, \citenamefont {Whittaker},
  \citenamefont {Lehnert},\ and\ \citenamefont
  {Simmonds}}]{teufel_sideband_2011}%
  \BibitemOpen
  \bibfield  {author} {\bibinfo {author} {\bibfnamefont {J.~D.}\ \bibnamefont
  {Teufel}}, \bibinfo {author} {\bibfnamefont {T.}~\bibnamefont {Donner}},
  \bibinfo {author} {\bibfnamefont {D.}~\bibnamefont {Li}}, \bibinfo {author}
  {\bibfnamefont {J.~W.}\ \bibnamefont {Harlow}}, \bibinfo {author}
  {\bibfnamefont {M.~S.}\ \bibnamefont {Allman}}, \bibinfo {author}
  {\bibfnamefont {K.}~\bibnamefont {Cicak}}, \bibinfo {author} {\bibfnamefont
  {A.~J.}\ \bibnamefont {Sirois}}, \bibinfo {author} {\bibfnamefont {J.~D.}\
  \bibnamefont {Whittaker}}, \bibinfo {author} {\bibfnamefont {K.~W.}\
  \bibnamefont {Lehnert}}, \ and\ \bibinfo {author} {\bibfnamefont {R.~W.}\
  \bibnamefont {Simmonds}},\ }\href@noop {} {\bibfield  {journal} {\bibinfo
  {journal} {Nature}\ }\textbf {\bibinfo {volume} {475}},\ \bibinfo {pages}
  {359} (\bibinfo {year} {2011}{\natexlab{b}})}\BibitemShut {NoStop}%
\bibitem [{\citenamefont {Castellanos-Beltran}\ \emph
  {et~al.}(2008)\citenamefont {Castellanos-Beltran}, \citenamefont {Irwin},
  \citenamefont {Hilton}, \citenamefont {Vale},\ and\ \citenamefont
  {Lehnert}}]{castellanos-beltran_amplification_2008}%
  \BibitemOpen
  \bibfield  {author} {\bibinfo {author} {\bibfnamefont {M.~A.}\ \bibnamefont
  {Castellanos-Beltran}}, \bibinfo {author} {\bibfnamefont {K.~D.}\
  \bibnamefont {Irwin}}, \bibinfo {author} {\bibfnamefont {G.~C.}\ \bibnamefont
  {Hilton}}, \bibinfo {author} {\bibfnamefont {L.~R.}\ \bibnamefont {Vale}}, \
  and\ \bibinfo {author} {\bibfnamefont {K.~W.}\ \bibnamefont {Lehnert}},\
  }\href@noop {} {\bibfield  {journal} {\bibinfo  {journal} {Nat Phys}\
  }\textbf {\bibinfo {volume} {4}},\ \bibinfo {pages} {929} (\bibinfo {year}
  {2008})}\BibitemShut {NoStop}%
\bibitem [{\citenamefont {Kimble}\ \emph {et~al.}(2001)\citenamefont {Kimble},
  \citenamefont {Levin}, \citenamefont {Matsko}, \citenamefont {Thorne},\ and\
  \citenamefont {Vyatchanin}}]{kimble_conversion_2001}%
  \BibitemOpen
  \bibfield  {author} {\bibinfo {author} {\bibfnamefont {H.~J.}\ \bibnamefont
  {Kimble}}, \bibinfo {author} {\bibfnamefont {Y.}~\bibnamefont {Levin}},
  \bibinfo {author} {\bibfnamefont {A.~B.}\ \bibnamefont {Matsko}}, \bibinfo
  {author} {\bibfnamefont {K.~S.}\ \bibnamefont {Thorne}}, \ and\ \bibinfo
  {author} {\bibfnamefont {S.~P.}\ \bibnamefont {Vyatchanin}},\ }\href@noop {}
  {\bibfield  {journal} {\bibinfo  {journal} {Phys. Rev. D}\ }\textbf {\bibinfo
  {volume} {65}},\ \bibinfo {pages} {022002} (\bibinfo {year}
  {2001})}\BibitemShut {NoStop}%
\bibitem [{\citenamefont {Demkowicz-Dobrzański}\ \emph
  {et~al.}(2013)\citenamefont {Demkowicz-Dobrzański}, \citenamefont
  {Banaszek},\ and\ \citenamefont
  {Schnabel}}]{demkowicz_dobrzanski_fundamental_2013}%
  \BibitemOpen
  \bibfield  {author} {\bibinfo {author} {\bibfnamefont {R.}~\bibnamefont
  {Demkowicz-Dobrzański}}, \bibinfo {author} {\bibfnamefont {K.}~\bibnamefont
  {Banaszek}}, \ and\ \bibinfo {author} {\bibfnamefont {R.}~\bibnamefont
  {Schnabel}},\ }\href@noop {} {\bibfield  {journal} {\bibinfo  {journal}
  {Phys. Rev. A}\ }\textbf {\bibinfo {volume} {88}},\ \bibinfo {pages} {041802}
  (\bibinfo {year} {2013})}\BibitemShut {NoStop}%
\bibitem [{\citenamefont {Leonhardt}(2005)}]{leonhart_2005}%
  \BibitemOpen
  \bibfield  {author} {\bibinfo {author} {\bibfnamefont {U.}~\bibnamefont
  {Leonhardt}},\ }\href@noop {} {\emph {\bibinfo {title} {Measuring the
  {Quantum} {State} of {Light}}}}\ (\bibinfo  {publisher} {Cambridge University
  Press},\ \bibinfo {address} {Cambridge, N.Y},\ \bibinfo {year}
  {2005})\BibitemShut {NoStop}%
\bibitem [{\citenamefont {Caves}\ \emph {et~al.}(1980)\citenamefont {Caves},
  \citenamefont {Thorne}, \citenamefont {Drever}, \citenamefont {Sandberg},\
  and\ \citenamefont {Zimmermann}}]{caves_measurement_1980}%
  \BibitemOpen
  \bibfield  {author} {\bibinfo {author} {\bibfnamefont {C.~M.}\ \bibnamefont
  {Caves}}, \bibinfo {author} {\bibfnamefont {K.~S.}\ \bibnamefont {Thorne}},
  \bibinfo {author} {\bibfnamefont {R.~W.~P.}\ \bibnamefont {Drever}}, \bibinfo
  {author} {\bibfnamefont {V.~D.}\ \bibnamefont {Sandberg}}, \ and\ \bibinfo
  {author} {\bibfnamefont {M.}~\bibnamefont {Zimmermann}},\ }\href@noop {}
  {\bibfield  {journal} {\bibinfo  {journal} {Rev. Mod. Phys.}\ }\textbf
  {\bibinfo {volume} {52}},\ \bibinfo {pages} {341} (\bibinfo {year}
  {1980})}\BibitemShut {NoStop}%
\bibitem [{\citenamefont {Arcizet}\ \emph {et~al.}(2006)\citenamefont
  {Arcizet}, \citenamefont {Briant}, \citenamefont {Heidmann},\ and\
  \citenamefont {Pinard}}]{arcizet_beating_2006}%
  \BibitemOpen
  \bibfield  {author} {\bibinfo {author} {\bibfnamefont {O.}~\bibnamefont
  {Arcizet}}, \bibinfo {author} {\bibfnamefont {T.}~\bibnamefont {Briant}},
  \bibinfo {author} {\bibfnamefont {A.}~\bibnamefont {Heidmann}}, \ and\
  \bibinfo {author} {\bibfnamefont {M.}~\bibnamefont {Pinard}},\ }\href@noop {}
  {\bibfield  {journal} {\bibinfo  {journal} {Phys. Rev. A}\ }\textbf {\bibinfo
  {volume} {73}},\ \bibinfo {pages} {033819} (\bibinfo {year}
  {2006})}\BibitemShut {NoStop}%
\bibitem [{\citenamefont {Kerdoncuff}\ \emph {et~al.}(2015)\citenamefont
  {Kerdoncuff}, \citenamefont {Hoff}, \citenamefont {Harris}, \citenamefont
  {Bowen},\ and\ \citenamefont
  {Andersen}}]{kerdoncuff_squeezing_enhanced_2015}%
  \BibitemOpen
  \bibfield  {author} {\bibinfo {author} {\bibfnamefont {H.}~\bibnamefont
  {Kerdoncuff}}, \bibinfo {author} {\bibfnamefont {U.~B.}\ \bibnamefont
  {Hoff}}, \bibinfo {author} {\bibfnamefont {G.~I.}\ \bibnamefont {Harris}},
  \bibinfo {author} {\bibfnamefont {W.~P.}\ \bibnamefont {Bowen}}, \ and\
  \bibinfo {author} {\bibfnamefont {U.~L.}\ \bibnamefont {Andersen}},\
  }\href@noop {} {\bibfield  {journal} {\bibinfo  {journal} {Annalen der
  Physik}\ }\textbf {\bibinfo {volume} {527}},\ \bibinfo {pages} {107}
  (\bibinfo {year} {2015})}\BibitemShut {NoStop}%
\bibitem [{\citenamefont {Braginsky}\ \emph {et~al.}(1980)\citenamefont
  {Braginsky}, \citenamefont {Vorontsov},\ and\ \citenamefont
  {Thorne}}]{braginsky_quantum_1980}%
  \BibitemOpen
  \bibfield  {author} {\bibinfo {author} {\bibfnamefont {V.~B.}\ \bibnamefont
  {Braginsky}}, \bibinfo {author} {\bibfnamefont {Y.~I.}\ \bibnamefont
  {Vorontsov}}, \ and\ \bibinfo {author} {\bibfnamefont {K.~S.}\ \bibnamefont
  {Thorne}},\ }\href@noop {} {\bibfield  {journal} {\bibinfo  {journal}
  {Science}\ }\textbf {\bibinfo {volume} {209}},\ \bibinfo {pages} {547}
  (\bibinfo {year} {1980})}\BibitemShut {NoStop}%
\bibitem [{\citenamefont {Huang}\ and\ \citenamefont
  {Agarwal}(2009)}]{huang_entangling_2009}%
  \BibitemOpen
  \bibfield  {author} {\bibinfo {author} {\bibfnamefont {S.}~\bibnamefont
  {Huang}}\ and\ \bibinfo {author} {\bibfnamefont {G.~S.}\ \bibnamefont
  {Agarwal}},\ }\href@noop {} {\bibfield  {journal} {\bibinfo  {journal} {New
  J. Phys.}\ }\textbf {\bibinfo {volume} {11}},\ \bibinfo {pages} {103044}
  (\bibinfo {year} {2009})}\BibitemShut {NoStop}%
\bibitem [{\citenamefont {Metelmann}\ and\ \citenamefont
  {Clerk}(2014)}]{metelmann_quantum-limited_2014}%
  \BibitemOpen
  \bibfield  {author} {\bibinfo {author} {\bibfnamefont {A.}~\bibnamefont
  {Metelmann}}\ and\ \bibinfo {author} {\bibfnamefont {A.}~\bibnamefont
  {Clerk}},\ }\href@noop {} {\bibfield  {journal} {\bibinfo  {journal} {Phys.
  Rev. Lett.}\ }\textbf {\bibinfo {volume} {112}},\ \bibinfo {pages} {133904}
  (\bibinfo {year} {2014})}\BibitemShut {NoStop}%
\end{thebibliography}%

\clearpage

\section{Supplementary Methods}
\subsection{Description of the Microwave Setup}
\begin{figure*}[htbp]
	\centering
	\includegraphics[width=15cm]{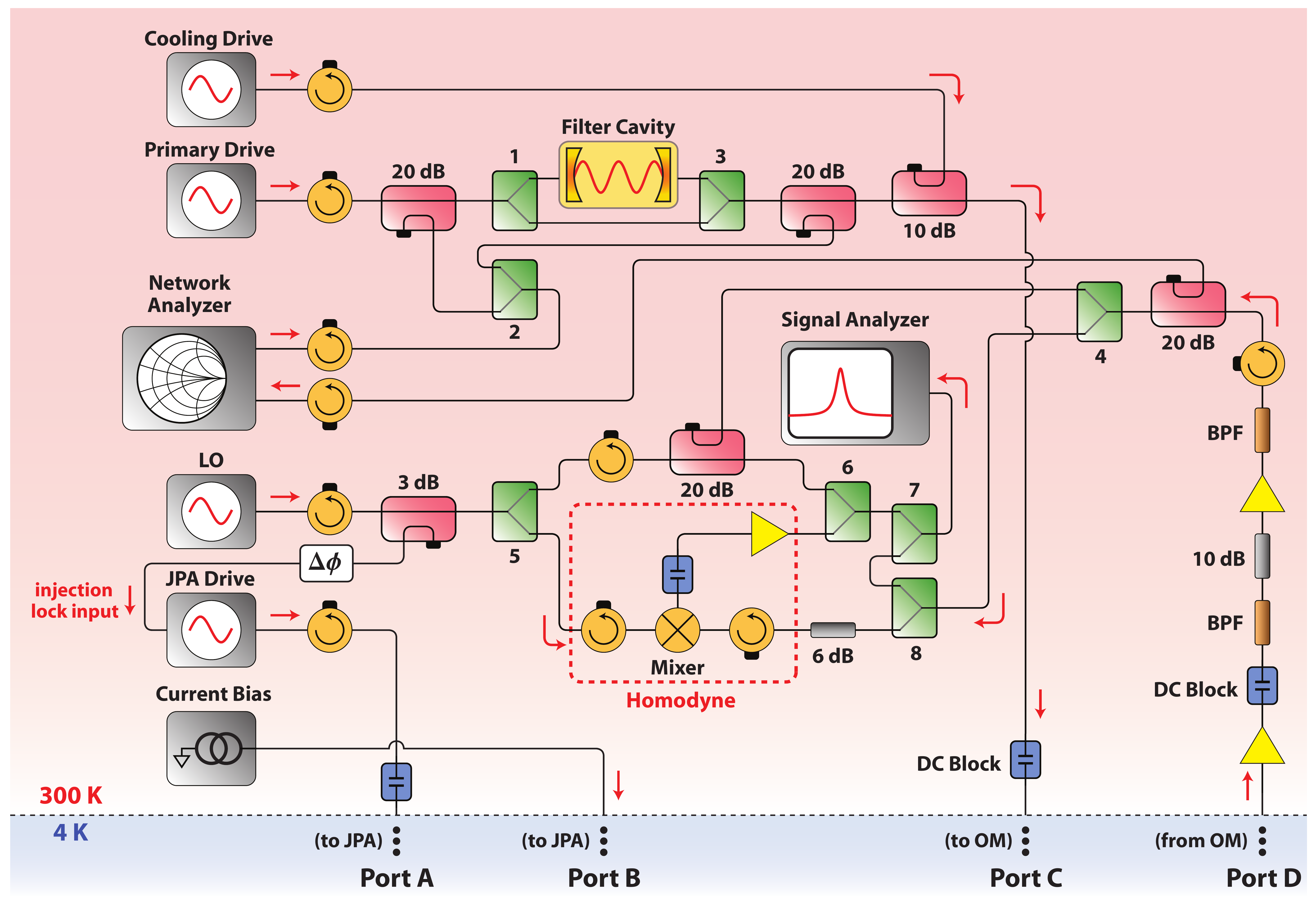}
	\caption{\label{fig:setup}
		Room temperature microwave setup.
		LO, local oscillator; BPF, 5.6--7.0 GHz bandpass filter; $\Delta\phi$, programmable phase shifter.
		The components shaded in green denote microwave switches, which have been numbered for reference.
		We lock the network analyzer, signal analyzer, microwave generators, and local oscillator (LO) to a 10 MHz reference provided by a $^{85}$Rb oven clock (not pictured).
		The frequency of the JPA drive was locked to twice the frequency of the primary drive by injection locking the internal doubler of the JPA microwave source.
		The ``primary" drive refers to the microwave drive tuned to the resonance frequency of the optomechanical cavity.
		The primary drive is filtered with a copper cavity to eliminate any excess noise (above the Johnson noise floor) at the mechanical sideband frequencies.
		The cooling drive was detuned by approximately 1.5 mechanical resonance frequencies to the red of the cavity resonance frequency (see Fig.~\ref{fig:noise}a), and its power was held constant to damp the mechanical mode to a factor of 10 above its intrinsic linewidth.
	}
\end{figure*}

\begin{figure}[htbp]
	\centering
	\includegraphics[width=\columnwidth]{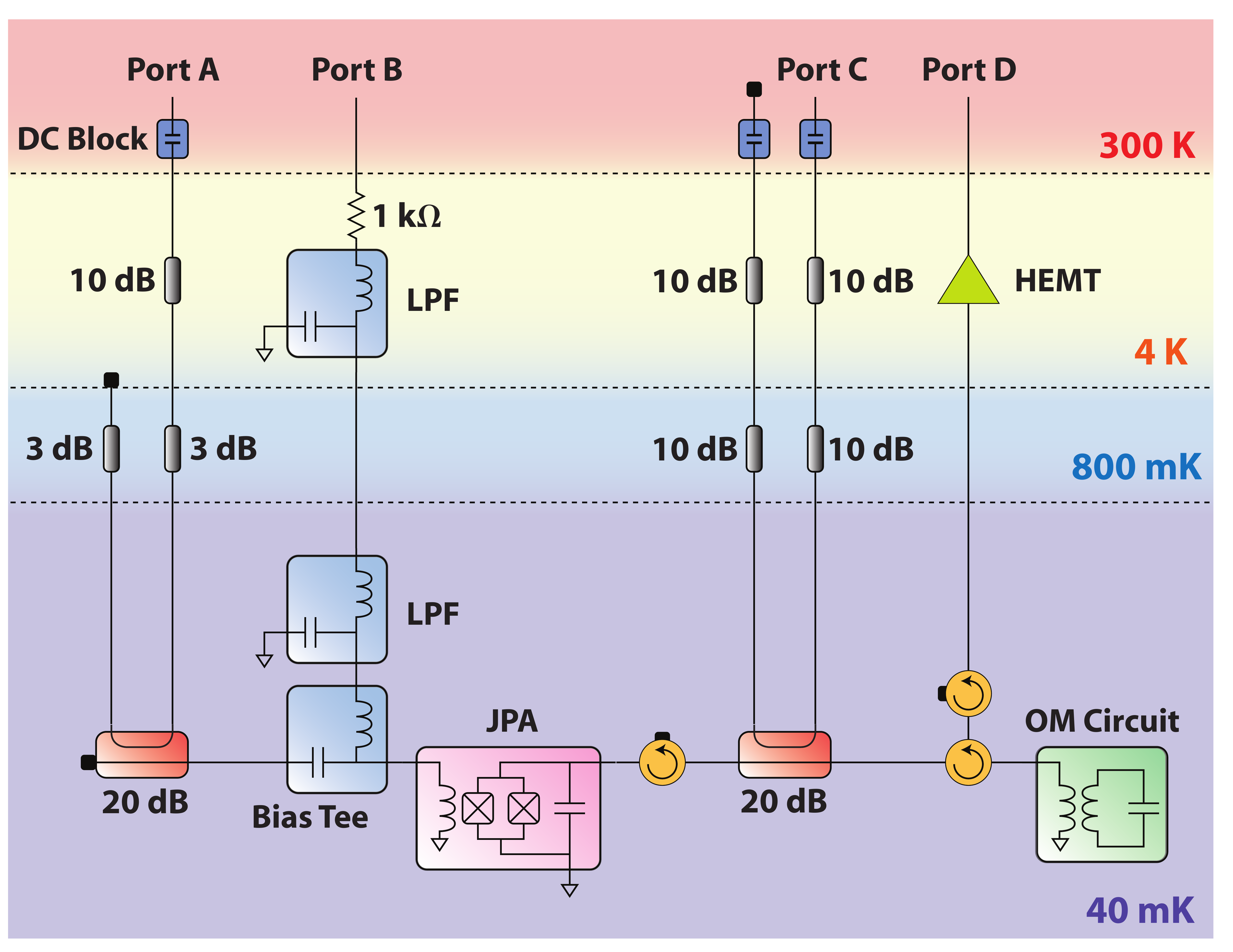}
	\caption{\label{fig:fridge}
		Circuit thermalization stages within the dilution cryostat.
	}
\end{figure}

Figure~\ref{fig:setup} details the assembly of microwave sources, receivers, and signal paths at room temperature that couple into and out of the dilution cryostat.
Figure~\ref{fig:fridge} complements Fig.~\ref{fig:setup} by detailing the thermalization of the drives within the cryostat.
The majority of the discussion in Fig.~\ref{fig:setup} is contained in the caption.
The salient features to note are:
\begin{enumerate}
	\item All microwave generators and receivers are frequency locked together using a combination of a master clock and master-slave injection locking.
	\item The ``primary" microwave drive to the optomechanical circuit is filtered at room temperature to reach the shot noise limit at base temperature at critical mechanical sideband frequencies.
	\item The microwave network is interconnected using programmable switches, enabling automated, reproducible toggling between calibrations (\textit{e.g.} probing the response of the filter cavity, calibrating phase offsets, \textit{etc}.) and data acquisition.
\end{enumerate}

\subsection{Calibration and Control of Drive Phases}
We describe the preparation of the relative phases of the microwave drives.
We are sensitive to the level of liquid helium in the 4K bath of the dilution refrigerator since any helium boil-off changes the effective microwave path lengths over time scales on the order of an hour.
This path length variation shifts the relative phase between the drives to the Josephson Parametric Amplifier (JPA) and optomechanical circuit, as well as the phase of the quadrature of the reflected drive being detected at room temperature.
Accordingly, we perform an initial calibration to account for these changing phases at the beginning of every experiment.
To do so, we turn off the drive to the JPA that generates the squeezing, toggling switches 4, 5, 6, and 7 (see Fig.~\ref{fig:setup}) so that the ``primary" drive returning from the cryostat interferes with the local oscillator (LO).
The beat note power is measured as the LO phase is varied to orient the measurement in the $X$-$Y$ quadrature plane of the returned drive.
We estimate an error in the phase selection to be approximately $\pm1^\circ$.
At the end of this calibration, we select the amplitude quadrature of the returned drive for detection and confirm that the power spectral density of the mechanical sideband is nulled to the noise floor (see Fig.~\ref{fig:qnd}).
Finally, the phase quadrature of the drive is selected and the mechanical sideband data (\textit{i.e.} the phase modulation of the returned drive) is recorded.

To prepare the desired manifold of displaced squeezed states, we must also be able to control the phase of the JPA drive ($\theta$) relative to the phase of the primary drive ($\phi$).
To meet this requirement, we pick off half of the LO power (which has been phase-referenced to the primary drive phase as explained in the previous paragraph) and send it into a programmable phase shifter.
The output of the phase shifter is then sent to the input LO port of a second microwave generator, which produces a doubled, phase-locked drive that is sent to the JPA.

\subsection{Further Data Acquisition Details}
We describe the measurement settings used to obtain the data presented in the manuscript.
To obtain the data in Figs.~2 and 3 of the main text, we tuned the JPA drive frequency to twice that of the ``primary" squeezed drive ($\omega_c/2\pi\approx6.78$~GHz).
The mechanical sideband was detected via homodyne measurement by toggling programmable switches 4--8 in Fig.~\ref{fig:setup} and tuning the LO to the same frequency as the primary drive.
Since increasing the drive power pulls the optomechanical cavity resonance frequency down, the primary drive frequency had to be adjusted as the drive power was increased.
We defined the optomechanical resonance frequency at high drive powers to be the frequency at which the drive yielded a consistent mechanical linewidth $\Gamma=2\pi\times200$~Hz (in the presence of the auxiliary cooling drive).
This search was executed before every experiment by sweeping the drive frequency at a given power, measuring the resulting linewidths, and interpolating the results to find $\Gamma=2\pi\times200$~Hz.
When the search was completed, we selected a measurement frequency span of approximately 300 kHz and a resolution bandwidth of 50 Hz, yielding 6000 independent points.
Each trace of the detected power spectral density was averaged 300 times before being analyzed.

To acquire the data in Fig.~4 of the main text, we introduced an intentional detuning between the drive frequency to the JPA and the drive frequency to the optomechanical cavity.
The primary drive frequency was again tuned to the cavity resonance frequency ($\omega_c$), but a slight detuning for the JPA drive frequency ($\omega_{\mathrm{JPA}}=2\omega_c+2\pi\times0.1$~Hz) was selected.
The detuning achieves a constant rate of rotation of the squeezing phase, $\theta$, relative to the phase of the coherent amplitude of the drive, $\phi$.
The power spectral density of the upper mechanical sideband was detected via heterodyne measurement by toggling switches 4, 7, and 8, sending the reflected drive frequency and its mechanical sidebands directly to a signal analyzer.
We selected a resolution bandwidth of 1 kHz in zero span, amounting to an integration of the total power spectral density over 5 mechanical linewidths.
We obtained a degree of averaging for each 200 point trace of the integrated power by selecting a video bandwidth of 2~Hz.
To build statistics, we performed 20 experiments at any given combination of measurement cooperativity and JPA drive strength.

\section{Supplementary Discussion}
\subsection{Ideality of the QND Measurement}
\begin{figure}[htbp]
	\centering
	\includegraphics[width=\columnwidth]{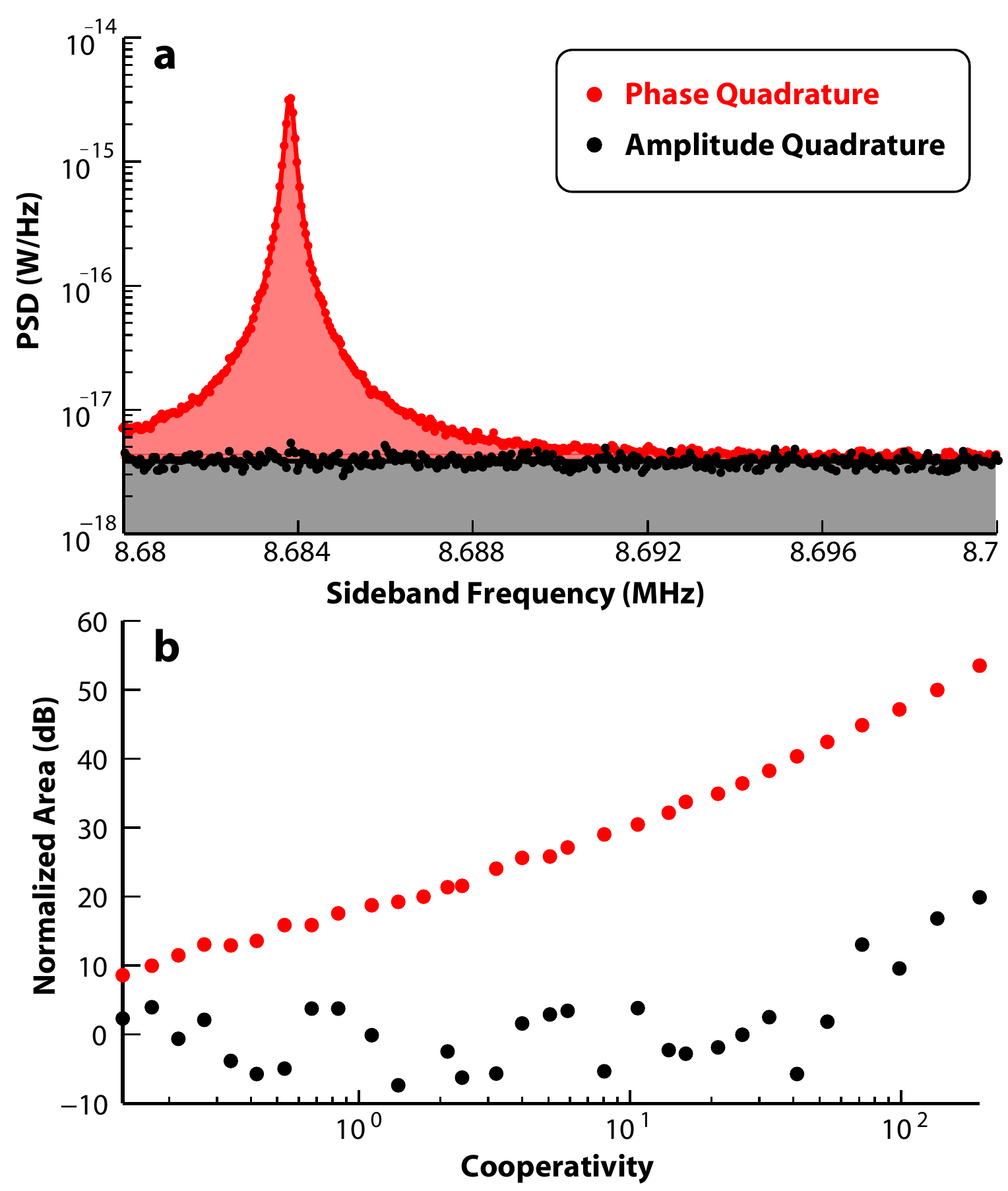}
	\caption{\label{fig:qnd}
		Characterizing the ideality of the quantum nondemolition (QND) measurement of the light field's amplitude quadrature.
		\textbf{a}, Detected power spectral densities near mechanical resonance of the amplitude (black) and phase (red) quadratures of the returned drive field (unsqueezed, measurement cooperativity of $C=50$).
		We see no detectable backaction on the amplitude quadrature despite the light field's interaction with the mechanical mode.
		On the other hand, the noise power of the light's phase quadrature, which is coupled to the mechanical motion, acts as an indirect probe of the light's amplitude fluctuations near mechanical resonance.
		\textbf{b}, Normalized areas of the Lorentzian line shapes near the mechanical resonance frequency for the light's amplitude and phase quadratures as a function of the drive strength.
		For the y-axis, 0~dB corresponds to the average measured power over the expected mechanical linewidth due to the fluctuations of the measurement noise floor.
		We do not observe any significant backaction on the amplitude quadrature until reaching cooperativities near 100.
	}
\end{figure}
We briefly comment on the ideality the quantum nondemolition (QND) measurement of the light field pursued in our experiments.
Here, the system observable of interest (the amplitude quadrature of the drive field, $\hat{X}_a$) couples to the phonon occupancy of the mechanical oscillator, which acts as the meter.
The motion of the mechanical oscillator is then re-encoded purely into the phase quadrature of the drive field.
Thus, we detect the phase quadrature of the returned drive in order to measure the drive's amplitude fluctuations.
The amplitude quadrature itself is unperturbed by measurement backaction and remains available for further measurements with the homodyne receiver.

The degree to which the QND observable suffers any backaction provides a useful characterization of the nonideality of the QND measurement process.
The effective measurement efficiency in Fig.~4c does much to demonstrate a lack of significant backaction on the amplitude quadrature since any added noise would appear as a degradation of the effective measurement efficiency, $\etaeff^{\mathrm{OM}}$.
It is nevertheless worthwhile to directly search for any detectable backaction on the light's amplitude quadrature after its interaction with the optomechanical cavity.
As Fig.~\ref{fig:qnd} illustrates, we do not observe any detectable excess amplitude noise above of the noise floor at measurement cooperativities below 50.
Any measurable excess noise on the light field's amplitude quadrature at higher measurement cooperativities is confined to be approximately 30~dB below the signal in the phase quadrature.
This level may be due to phase noise among the drives and local oscillator or due to imprecise detuning of the primary drive from cavity resonance.

\subsection{Model of the Drive}
\begin{figure}[htbp]
\centering
\includegraphics[width=\columnwidth]{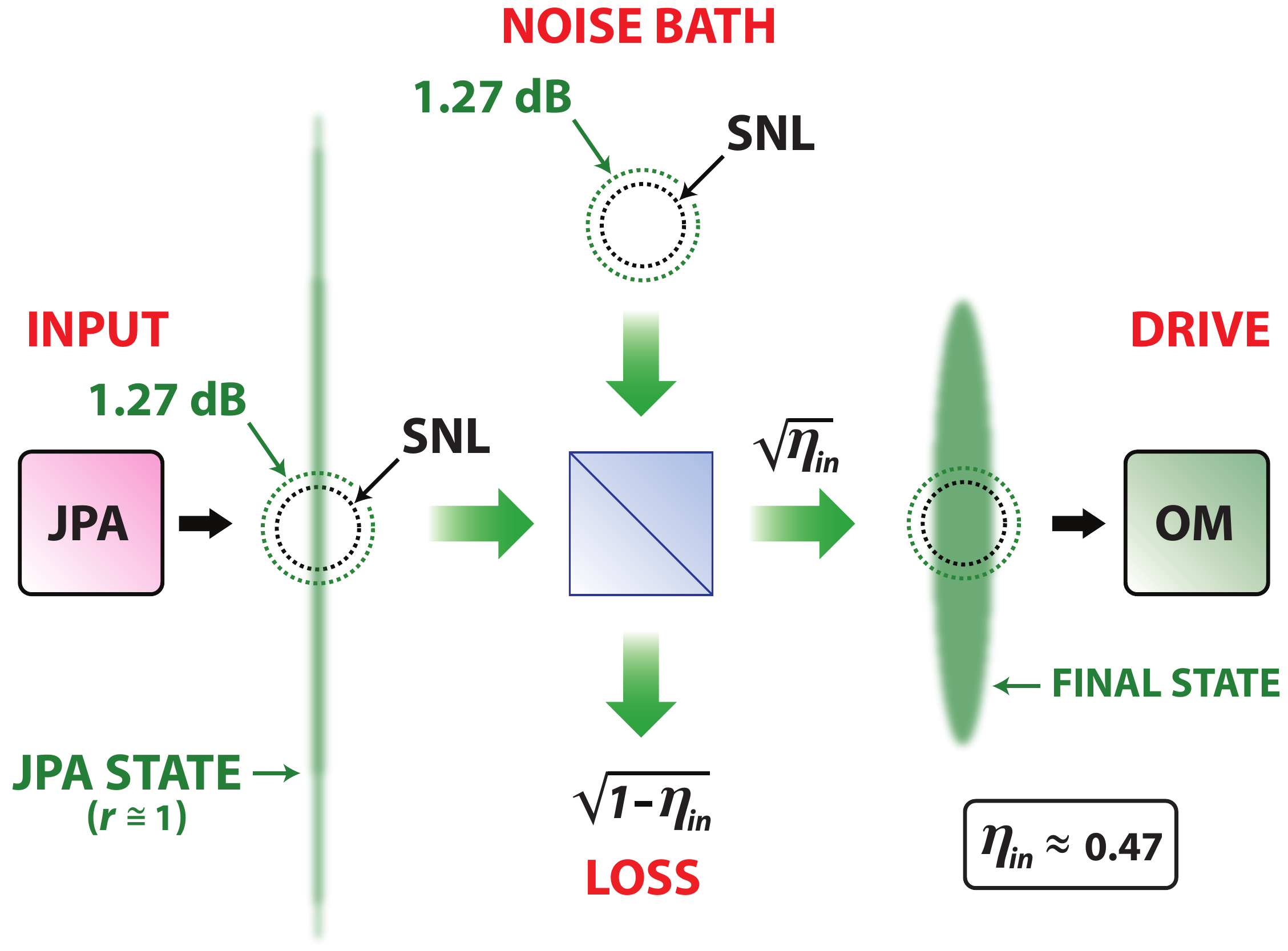}
\caption{\label{fig:model}
	Model of the squeezed state of the drive.
	For clarity, the displacement of the noise ellipse from the origin is not pictured since it does not affect the quadrature variances (which define the entropy of the state and, by extension, the ultimate noise performance of the detection).
	The interrogating field can be thought of as a squeezed thermal state (1.27 dB of excess noise above the SNL, which drives the cavity thermal occupancy to $n_c=0.17$) mixing with a thermal noise bath at equilibrium with the temperature of the input.
	Thus, the impurity of the drive results from both the loss (at the beamsplitter in this model) and the thermal occupancy of the inputs.
	We infer an effective beamsplitter transmittance $\etain\approx47\pm2\%$ from the data.
}
\end{figure}

We describe our model for the displaced squeezed states that were used to obtain data in Figs.~2--4 of the main text.
In principle, one might expect to treat the drive field as a pure displaced squeezed state subject to loss with vacuum.
As explained in the manuscript, however, we consistently observed a small degree of excess backaction in the presence of an unsqueezed drive.
Moreover, this degree excess backaction was preserved in the presence of both amplitude-- and phase--squeezed states alike, and it did not depend on the applied drive power.
This behavior is consistent with a squeezed thermal state subject to loss, so we modeled the state as such.
The loss model assumes that the drive is mixed with an additional noise bath (via a beamsplitter of transmittance $\etain\approx47\pm2\%$, see Fig.~\ref{fig:model}) at thermal equilibrium with the bath being squeezed.
We characterize the temperature of these baths by an excess thermal occupancy of the optomechanical cavity, $\mathrm{Tr}(\hat{\rho}_a\hat{a}^\dagger\hat{a})\equiv n_c=0.17$.
Accordingly, the variance of the generalized quadrature $\hat{X}_\theta\equiv\hat{X}\cos\theta+\hat{Y}\sin\theta$ of the squeezed state prepared by the JPA goes as
\begin{align}
&\frac{\braket{(\Delta\hat{X}_\theta)^2}}{\mathrm{SNL}}=\\
&(1+2n_c)\times\bigg(1-\etain+\etain(\cosh2r-\cos\theta\sinh2r)\bigg)\nonumber.
\end{align}

We verify the validity of this model by performing a variety of related measurements.
Given an individual imprecision-backaction curve (as portrayed in Fig.~3a of the main text), it is possible to neglect the effects of a small amount of excess noise on the drive by adjusting the measurement cooperativity and detection efficiency together to fit the data.
Accordingly, we independently calibrated the power-dependence of the measurement cooperativity after the devices were cooled down by measuring the optical damping achieved by a red-detuned drive.
Moreover, at the beginning of each experiment, we retrieve the cooperativity from fitting the network scattering response ($S_{11}$) of the optomechanical circuit near the upper mechanical sideband while the cavity is pumped at its resonance frequency (often referred to as the ``driven" response of the circuit).

With the cooperativity determined, any excess backaction is straightforward to measure.
Furthermore, knowing the cooperativity constrains the detection efficiency via the slope of the ``imprecision" side of the imprecision-backaction curve.
With the detection efficiency constrained, we are able to extract the squeezing parameter associated with a given JPA drive power by a simple homodyne detection of the JPA output (in the absence of any coherent drive to the optomechanical cavity).

\begin{figure}[htbp]
	\centering
	\includegraphics[width=\columnwidth]{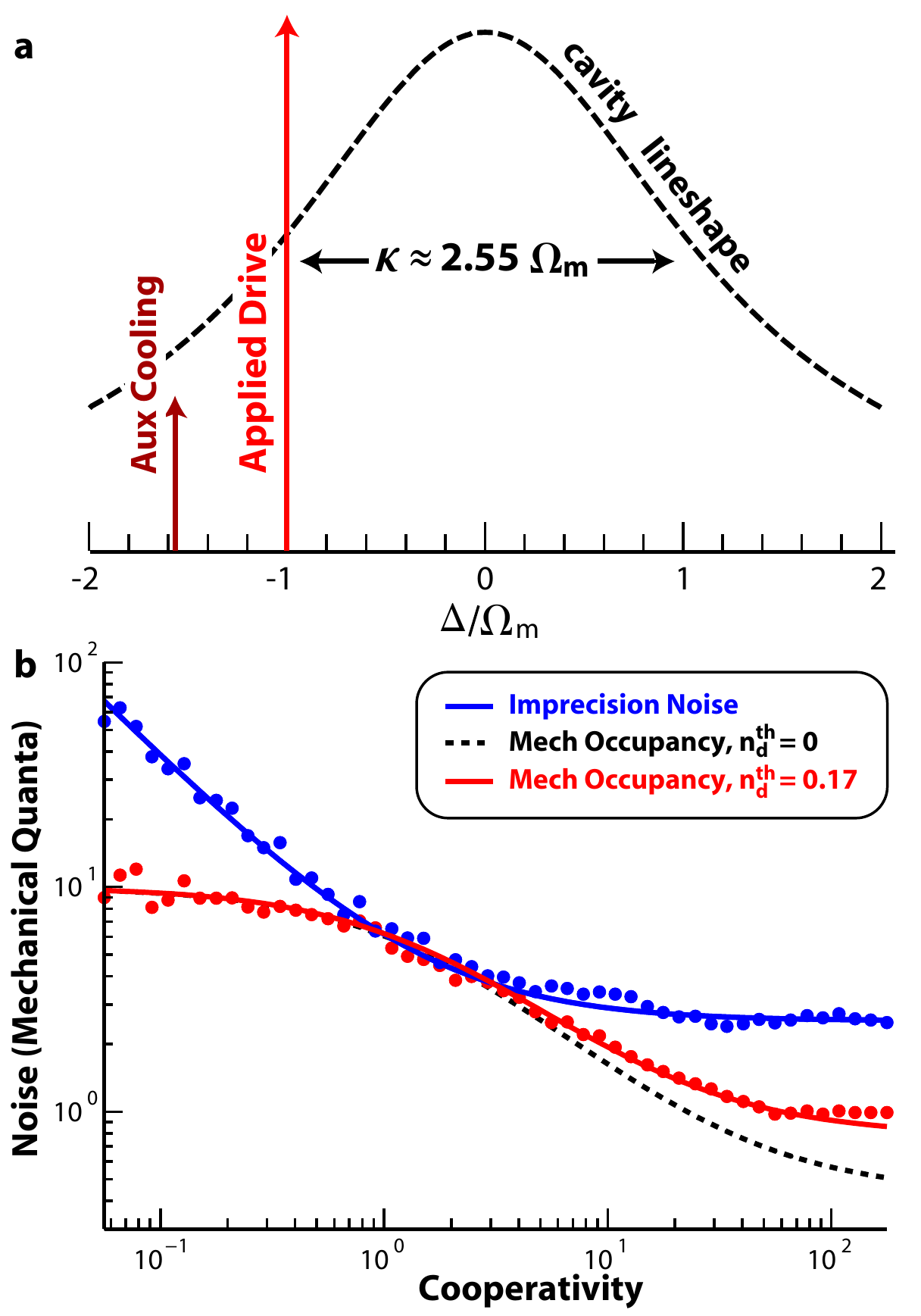}
	\caption{\label{fig:noise}
		Confirmation of the postulated level of excess drive noise via sideband cooling.
		\textbf{a}, Schematic of the drive detuning used for the cooling experiment.
		The power and detuning of the ``Aux" cooling drive are identical to those of the constant damping drive used in the experiments.
		We therefore achieve the same initial phonon occupancy as the experiments presented in the manuscript (10 quanta) and the same total mechanical linewidth (200 Hz).
		\textbf{b}, Imprecision noise ($\nimp$) and thermal phonon occupancy $\nf$ inferred by a measurement of the upper mechanical sideband.
		We observe a deviation from the ideal cooling behavior of a shot-noise-limited drive (\textit{i.e.} a pure coherent state), which is indicated by the black dashed line.
		The theory curve plotted in red assumes a cavity thermal occupancy of $n_c=0.17$, which corresponds to the model used to construct the imprecision/backaction data.
	}
\end{figure}

As a final check of the validity of all experimental parameters, we performed a simple cooling experiment with the same damping drive power and detuning as the experiments detailed in the manuscript.
In addition to this damping drive, we ramped a separate microwave drive tuned to the red of the cavity resonance frequency by the mechanical resonance frequency as detailed in Fig.~\ref{fig:noise}a.
The data confirm that the cooling performance achieved by the unsqueezed primary drive is consistent with a thermal cavity occupancy of $\nc=0.17$.

\section{Supplementary Equations}
\subsection{Effects of the Auxiliary Cooling Tone}
We introduced an auxiliary red-detuned drive (apart from the primary squeezed drive) to sideband cool the mechanics to a lower initial mode temperature and to stabilize the mechanics with an optical damping force.
This cooling drive introduces additional noise effects in the presence of the squeezing of the primary drive.
What is critical to notice is that the quantum correlations associated with the squeezing are only correlated in the frame of the primary drive field.
The upper and lower sideband fluctuations of the primary drive are correlated with each other in the presence of the squeezing, but the upper and lower sidebands relative to the damping drive are not.
Thus, from the perspective of the damping drive, the introduction of squeezed vacuum effectively acts as a source of excess quadrature noise above the shot noise limit (SNL). 

In the presence of a strong drive cooperativity, any thermal impurity of the cavity drive field couples to the equilibrium mechanical state in the form of excess backaction.
When sideband cooling the mechanics in the presence of a cavity thermal occupancy $n_c$, the equilibrium phonon occupancy of a perfectly overcoupled optomechanical cavity takes the form
\begin{equation}
\label{eq:finaloc}
\nf = \frac{\Gamma_m\times\nth + \Gamma_-^{\mathrm{cool}}\times(1+n_c) + \Gamma_+^{\mathrm{cool}}\times n_c}{\Gamma_m - \Gamma_-^{\mathrm{cool}} + \Gamma_+^{\mathrm{cool}}},
\end{equation}
where $\Gamma_{\pm}^{\mathrm{cool}}$ denote the anti-Stokes and Stokes scattering rates of the cooling drive.
Thus, if the strength of the cooling drive is limited such that
\begin{equation}
\label{eq:condition}
\sinh^2r\times\Gamma_+^{\mathrm{cool}}\ll\nth\times\Gamma_m,
\end{equation}
the equilibrium phonon occupancy, $\nf$, will be dominated by the incoherent noise of the thermal bath and not by the presence of the squeezing.
In our experiments, we select a cooling drive strength such that
\begin{equation}
\Gamma_+^{\mathrm{cool}}\approx 230~\mathrm{Hz}
\end{equation}
and squeezing parameters such that
\begin{equation}
\sinh^2r\le 1.5,
\end{equation}
which satisfies Eq.~\ref{eq:condition}.
We therefore ignore these higher order noise effects and treat the cooling drive as an ideal sideband cooling drive.

\subsection{Efficiency of the Mechanical Detector}
Measuring the values of $\etadetom$ in Fig.~4b of the main text (shaded red) consisted of two steps:
\begin{enumerate}
	\item Measuring the \textit{total} effective detection efficiency $\etaeff$ of the squeezing measurement
	\item Accounting for the contribution to $\etaeff$ due to microwave loss to obtain $\etaeff^{\mathrm{OM}}$
\end{enumerate}
We define the effective detection efficiency $\etaeff$ by applying the fit
\begin{align}
&\frac{\braket{(\Delta\hat{X}_a)^2}}{\mathrm{SNL}}=\\
&1-\etaeff+\etaeff\times\bigg(\cosh 2r -\cos\bigg(\frac{\phi+\theta}{2}\bigg)\sinh 2r\bigg)\nonumber
\end{align}
to the data obtained in Fig.~4a of the main text.
After obtaining $\etaeff$, we divide by the microwave transmittance $\etain=47\%$ between the JPA and the optomechanical circuit (see Fig.~\ref{fig:model}) to obtain the measured values of $\etadetom$.
The calculation of the band of expected values for the optomechanical detection efficiency (blue shaded region in Fig.~4b) numerically accounted for a variety of experimental parameters, including the finite integration bandwidth of the mechanical sideband's power spectral density and the excess background noise due to the squeezing.

In the main text, we invoked a more idealized expression for the effective detection efficiency that only depends on the mechanical properties, which we now discuss.
We first observe that the amplitude variance of any impure state of the drive, $[(\Delta\hat{X}_a)^2]\equiv\mathrm{Tr}(\hat{\rho}_a(\Delta\hat{X}_a)^2)$, is transduced to a measured mechanical phonon occupancy $\nf$ according to
\begin{equation}
\label{eq:occ}
\nf=\nth+\nimp+\frac{4C}{1+4(\tfrac{\Omega_m}{\kappa})^2}\times[(\Delta\hat{X}_a)^2].
\end{equation}
The constants and their experimental values used here are listed in Table~\ref{table:constants}.
If we divide the final mechanical occupancy of Eq.~\ref{eq:occ} by the mechanical occupancy in the presence of a coherent state drive at the same cooperativity (the effective ``shot noise limit" of the measurement), we can associate an optomechanical detection efficiency $\etadetom$ according to
\begin{align}
\label{eq:etadetmap}
& 1-\etadetom+\etadetom[(\Delta\hat{X})^2]=\\
&\frac{\nth+\nimp+4C(1+(\tfrac{\Omega_m}{\kappa})^2)^{-1}\times[(\Delta\hat{X})^2]}{\nth+\nimp+\nba^{\mathrm{coh}}}\nonumber.
\end{align}
Here $\nba^{\mathrm{coh}}=C\times(1+4(\tfrac{\Omega_m}{\kappa})^2)^{-1}$ denotes the mechanical occupancy due to backaction in the presence of a coherent drive.
By inspection of Eq.~\ref{eq:etadetmap}, we conclude that
\begin{equation}
\label{eq:etaom}
\etadetom = \bigg(1+\bigg(\frac{\nth+\nimp}{\nba^{\mathrm{coh}}}\bigg)\bigg)^{-1}.
\end{equation}
At sufficiently strong drives, $\nimp\ll\nth$, which allows us to drop $\nimp$ from Eq.~\ref{eq:etaom}.
By noting that $\nba^{\mathrm{coh}}\times\Gamma=\Gamma_{\mathrm{scatter}}$, we arrive at Eq.~4 of the main text.

\section{Supplementary Tables}
 \begin{table}[htbp]
 \caption{\label{table:constants}Summary of important experimental parameters.}
 \begin{tabular}{|l|c|c|}
		\hline
		\textbf{Parameter} & \textbf{Symbol} & \textbf{Value} \\
		\hline
		Cavity resonance frequency & $\omega_c$ & $2\pi\times$~6.89 GHz \\
		\hline
		Mechanical resonance frequency & $\Omega_m$ & $2\pi\times$~8.68 MHz \\
		\hline
		Intrinsic mechanical linewidth & $\Gamma_m$ & $2\pi\times$~22 Hz \\
		\hline
		Total mechanical linewidth & $\Gamma$ & $2\pi\times$~200 Hz \\
		\hline
		Total cavity linewidth & $\kappa$ & $2\pi\times$~22.2 MHz \\
		\hline
		Vacuum optomechanical  coupling rate & $g_0$ & $2\pi\times$~170 Hz \\
		\hline
		Measurement cooperativity & $C$ & $4g_0^2|\alpha|^2/\kappa\Gamma$ \\
		\hline
		Microwave transmittance (JPA to OM) & $\etain$ & 47\% \\
		\hline
		Drive excess noise (thermal occupancy) & $n_c$ & 0.17 \\
		\hline
		System noise temperature & $T_N$ & 5.5~K \\
		\hline
		Effective homodyne detection efficiency & $\eta_{\mathrm{det}}$ & 3\% \\
		\hline
		Base temperature of dilution cryostat & $T$ & 40 mK \\
		\hline
		Damped thermal phonon occupancy & $\nth$ & 10 \\
		\hline
 \end{tabular}
 \end{table}
 
\end{document}